\documentclass[12pt, a4paper]{article}
\pdfoutput=1
\usepackage{amsmath, amssymb,slashed,url,bm,upgreek}

\usepackage [T1]{fontenc}
\usepackage {calligra}
\usepackage [T1]{pbsi}
\usepackage[colorlinks=true,linkcolor=blue]{hyperref}
\usepackage{amsmath,amssymb,amsfonts} 
\usepackage{graphicx}                 
\usepackage{color}                    
\usepackage{hyperref}                 

\newcommand{\be}{\begin{equation}}
\newcommand{\ee}{\end{equation}}
\newcommand{\ba}{\begin{aligned}}
\newcommand{\ea}{\end{aligned}}
\newcommand{\bea}{\begin{eqnarray}}
\newcommand{\eea}{\end{eqnarray}}

\def\sech{\operatorname{sech}}
\def\tr{\operatorname{tr}}

\def\eps{{\epsilon}}

\def\rv{u}

\def\r{\rangle}

\def\1{\mathbf{1}}
\def\0{|\1\r}

\newcommand{\rme}{{\rm e}}
\newcommand{\rmi}{{\rm i}}
\newcommand{\rmd}{{\rm d}}

\begin{document}

\begin{flushright}
YITP-SB-14-13
\end{flushright}

\begin{center}
\vspace{1.5cm} { \Large {\bf 
Linear Resistivity 
from Non-Abelian Black Holes
}}
\vspace{1.2cm}

Christopher P.~Herzog, Kuo-Wei Huang, Ricardo Vaz

\vspace{1.2cm}

{\it C.~N.~Yang Institute for Theoretical Physics\\
Department of Physics and Astronomy \\
Stony Brook University, Stony Brook, NY  11794}

\vspace{1.2cm}

\end{center}

\begin{abstract}
\noindent
Starting with the holographic p-wave superconductor, we show how to obtain a finite DC conductivity through a non-abelian gauge transformation.  The translational symmetry is preserved. 
We obtain phenomenological similarities with high temperature cuprate superconductors. Our results suggest that a lattice or impurities are not essential to produce a finite DC resistivity with a linear temperature dependence.  An analogous field theory calculation for free fermions, presented in the appendix, indicates our results may be a special feature of strong interactions.
\end{abstract}

\pagebreak
\setcounter{page}{1}
\tableofcontents
\fontsize{11pt}{14pt}\selectfont
\section{Introduction}

Uncovering the mechanisms of high temperature superconductivity \cite{1986} has been one of the great challenges of theoretical and experimental physics.  
Strong interactions are believed to play an important role, rendering the conventional approach due to Bardeen, Cooper, and Schrieffer (BCS) \cite{BCS} inadequate.
An effective way to capture universal properties of high temperature superconductors is certainly of great interest.
The AdS/CFT correspondence \cite{Maldacena:1997re, Witten:1998qj, Gubser:1998bc}, which works by mapping certain strongly interacting field theories to classical theories of gravity, has proved to be a powerful tool for understanding strong interactions more generally.  Optimistically, one might hope that AdS/CFT might be able to provide some hints of the mechanisms underlying high temperature superconductivity.

The first holographic superconductors were constructed in refs.\  \cite{Gubser:2008px, Hartnoll:2008vx, Hartnoll:2008kx} where the black hole develops scalar hair at the phase transition. In refs.\ \cite{Gubser:2008zu, Gubser:2008wv, Roberts:2008ns}, the scalar was omitted and a non-abelian SU(2) gauge field was introduced, and the order parameter for the phase transition is the set of nonabelian global SU(2) currents. This non-abelian model provided a connection to a p-wave superconductor. 
While the initial p-wave papers focused on systems with two spatial dimensions, some corresponding analytic results were obtained for three spatial dimensions in ref.\ \cite{Herzog:2009ci} based on an analytic solution of the zero mode for the phase transition for an SU(2) gauge field in $AdS_5$ first observed in ref.\ \cite{Basu:2008bh}.
As recognized in ref.\ \cite{Hartnoll:2008vx}, these holographic systems are perhaps more accurately described as superfluids because the U(1) that would be associated with the photon is treated as a global symmetry.  But for many questions, the distinction may not be that important.

One of the most distinctive features of high-temperature cuprate superconductors is the linear temperature dependence of the DC electrical resistivity at optimal doping. However, a na\"{i}ve search for this effect in a dual gravitational model meets an immediate difficulty: The translation invariance in a gravitational model
implies momentum conservation. The charged particles then cannot dissipate their
momentum and the conductivity is infinite in the DC limit. 
Two obvious ways of breaking translation invariance, which real world materials take advantage of, are impurities and a lattice.
In an AdS/CFT context, enormous effort has recently gone into adding impurities and a lattice to gravity models.  
(See refs.\ \cite{Hartnoll:2007ih,Kachru:2009xf} for early papers on the subject.)\footnote{
A finite DC limit can also be achieved holographically by decoupling the charge current from the momentum current, for example by setting the total charge to zero \cite{Herzog:2007ij} or by holding the background metric fixed in a ``probe limit'' \cite{Hartnoll:2008vx,Karch:2007pd}.
}

In this paper, we present a new method, employing a non-abelian gauge transformation, 
that allows us to obtain a finite DC conductivity without breaking translation symmetry.
Moreover, we find the DC resistivity has a linear temperature dependence close to the superconducting phase transition.
Our most interesting result is the DC resistivity plot in Fig.\ \ref{rhoofT}. 
Our results suggest that a lattice or impurities are not necessary to produce a
finite DC resistivity with a linear temperature dependence.

The organization of the paper is as follows: In Sec.\ 2, we start from an action that is holographically dual to a p-wave superconductor. We next discuss the phase transitions for these systems with two and three spatial dimensions.  In Sec.\ 3, we compute the DC conductivity numerically as a function of temperature, using our non-abelian gauge transformation method. 
Motivated by the fact that the cuprate superconductors are layered materials, we will focus our numerical results on the $AdS_4$ model.
In Sec 4, we use the same idea to obtain an analytic form of a finite DC conductivity in $AdS_5$  using the solution given in \cite{Basu:2008bh}. We discuss some generalizations in the last section.
An appendix discusses the DC conductivity calculation for free fermions transforming under a global SU(2).  

\section{The p-wave Holographic Superconductor  }
\subsection{Action}
Our p-wave holographic superconductor has a dual description via the following gravitational action for a non-abelian gauge field $F_{\mu \nu}^a$ 
with a cosmological constant $\Lambda$:
\be
S = \frac{1}{2 \kappa^2} \int d^{d+1} x \sqrt{-g} \left( R - 2 \Lambda \right) - \frac{1}{4 e^2} \int d^{d+1} x \sqrt{-g} \, F^a_{\mu \nu} F^{a \mu\nu} \ .
\label{action}
\ee
For the moment, we keep $d$ arbitrary as we will study both the $d=3$ and $d=4$ cases.
Our gauge field is the curvature of the connection $A_\mu^a$:
\be
F_{\mu\nu}^a = \partial_\mu A_\nu^a - \partial_\nu A_\mu^a + {f^a}_{bc} A_\mu^b A_\nu^c \ ,
\ee
where ${f^a}_{bc}$ are the structure constants for our Lie algebra ${\mathfrak g}$ with generators $T_a$ such that $[T_a, T_b] = \rmi {f_{ab}}^c T_c$.  We will take ${\mathfrak g} =  {\rm su(2)}$ where $T_a = \sigma_a / 2$, $\sigma_a$ are the Pauli spin matrices, and the structure constants are $f_{abc} = \eps_{abc}$.  (The indices $a, b,c, \ldots$ are raised and lowered with $\delta^a_b$.)

The equations of motion for the gauge field that follow from this action (\ref{action}) are 
$D_\mu F^{a \mu\nu}= 0$ which can be expanded as
\be
\nabla_\mu F^{a \mu\nu} + {f^a}_{bc} A_\mu^b F^{c \mu\nu} = 0 \ .
\label{gaugefieldeom}
\ee
Einstein's equations can be written
\be
G_{\mu\nu} \equiv \frac{1}{2 \kappa^2} \left(  R_{\mu\nu} + \left( \Lambda - \frac{1}{2} R \right) g_{\mu\nu} \right) 
- \frac{1}{4 e^2} \left( 2
F^a_{\lambda\mu} {F^{a\lambda}}_\nu - \frac{1}{2} F^a_{\lambda\rho} F^{a \lambda\rho} g_{\mu\nu}  \right) =0 \ .
\ee

One well known solution to these equations in the case of a negative cosmological constant, $\Lambda= - d (d-1)/2 L^2$, is a Reissner-Nordstrom black hole with anti-de Sitter space asymptotics.  This solution describes the normal phase of the holographic p-wave superconductor.  The only nonzero component of the vector potential 
is\footnote{%
 The notation $\rho$ is mean to evoke a charge density.  The actual charge density according to the AdS/CFT dictionary would be
 $\tilde \rho = -(d-2) \rho / e^2$.
}
\be
A_t^3 \equiv \phi(u)  =   \mu + \rho \rv^{d-2} \ .
\label{Atansatz}
\ee
Thus we are using only a U(1) subgroup of the full SU(2) gauge symmetry; this black hole solution
requires only an abelian gauge symmetry.  The line element for this black hole solution has the form
\be
\frac{ds^2}{L^2} = \frac{-f(\rv)dt^2 + d \vec x^2}{\rv^2 } +  \frac{d\rv^2}{\rv^2 f(\rv)} \ 
\ee
where the warp factor is
\be
f(\rv) = 1 + Q^2 \left(\frac{\rv}{\rv_h}\right)^{2d-2} - \left(1+Q^2 \right) \left(\frac{\rv}{\rv_h} \right)^d \,
\label{warpfactor}
\ee
and the charge $Q$ has been defined as
\be
Q \equiv \lambda \rho \rv_h^{d-1} \sqrt{ \frac{d-2}{d-1}} \ ,
\ee
where we have defined the dimensionless parameter 
\be
\lambda \equiv \frac{\kappa}{e L} \,, 
\ee
controlling the back reaction on the metric.
The horizon is located at $\rv=\rv_h$, and the Hawking temperature is
\be
\label{HawkingT}
T_H = \frac{d - (d-2) Q^2}{4 \pi u_h} \ .
\ee
%
Our gauge potential (\ref{Atansatz}) is well defined globally, at both the horizon and the boundary, provided 
\be
\rho = - \frac{\mu}{\rv_h^{d-2}} \ .
\ee
\subsection{Superfluid Solution and Phase Diagram }
Increasing the chemical potential $\mu$ or equivalently decreasing the Hawking temperature, this black hole is well known  \cite{Gubser:2008wv,Roberts:2008ns} to undergo a phase transition to a state (dual to the superconducting state) with a nontrivial profile for
\be
A_x^1 \equiv w(u) \ .
\ee 
We need to reconstruct these results here as we will be exploring the conductivity close to the phase transition line.
Depending on the value of $\lambda$, this phase transition can be either first or second order \cite{Ammon:2009xh,Gubser:2010dm}.  
In the case of a second order phase transition, the location is given by the existence of a nontrivial zero mode solution for $w$ with regular boundary conditions at the horizon, $w(u_h) < \infty$, and  Dirichlet boundary conditions at the conformal boundary, $w(0) = 0$. 
The differential equation that must be solved is
\[
\rv^{d-3} f ( f \rv^{3-d} w')' = - \phi^2 w \ ,
\]
where $\phi$ is given by (\ref{Atansatz}), $f$ by (\ref{warpfactor}), and $f' \equiv \partial_u f$.

To find the location of the first order phase transition, we need to work harder and find a numerical solution for the condensed phase.  Following \cite{Ammon:2009xh}, we choose a metric ansatz
\be
\frac{ds^2}{L^2} = \frac{1}{\rv^2} \left( -f(u) s(u)^2 \, dt^2 + \frac{dx^2}{g(u)^{2(d-2)}} + g(u)^2 d \vec y^2 \right)  + \frac{d \rv^2}{f(u)} \ .
\label{metric}
\ee
We will be interested in $d=3$ or $d=4$ dimensions, and the relevant differential equations can be expressed by
\bea \label{eqsuperfluid}
u^{d-3} s \left( \frac{ \phi'}{u^{d-3}s} \right)' &=& \frac{g^{2(d-2)} w^2}{f} \phi\ , \nonumber \\
\frac{u^{d-3}}{ g^{2(d-2)}} \left( \frac{s g^{2(d-2)} f w'}{u^{d-3}} \right)' &=& - \frac{\phi^2}{f s} w\ , \nonumber\\
\frac{u^d}{s} \left( \frac{s f}{u^d} \right)' &=& \frac{\lambda^2}{d-1} \frac{ u^3 (\phi')^2}{ s^2} - \frac{d}{u} \ , \\
s' &=&- (d-2) \frac{(g')^2 u s }{g^2} - \frac{\lambda^2 u^3 g^{2(d-2)}}{d-1} \left( (w')^2  s+ \frac{ w^2 \phi^2}{ f^2 s}\right) \ , \nonumber\\
(d-1) \left( g'' - \frac{(g')^2}{g} \right)&=& -\lambda^2 u^2 g^{2d-3} \left( \frac{ w^2 \phi^2}{f^2 s^2} - (w')^2 \right)
- (d-1)\frac{g'}{u} \left( 1 + \frac{\lambda^2 u^4(\phi')^2}{(d-1)f s^2} - \frac{d}{f} \right) \  . \nonumber
\eea

To find the superfluid phase, we require these differential equations to have the following $u=0$ expansions in $d=3$:
\begin{eqnarray}
\label{bdry3begin}
\phi &=& \mu + \rho u + \frac{w_1^2 \mu}{12} u^4 + O(u^5) \ , \\
w &=& w_1 u - \frac{w_1 \mu^2}{6} u^3 + O(u^4) \ , \\
s &=& 1 - \frac{w_1^2 \lambda^2}{8} u^4 + O(u^6) \ , \\
g &=& 1 + g_3 u^3 + \frac{w_1^2 \lambda^2}{8} u^4 +O(u^6) \ , \\
f &=& 1 + f_3 u^3 + \frac{1}{2} \lambda^2 ( w_1^2 +  \rho^2) u^4 + O(u^6) \ ;
\label{bdry3}
\end{eqnarray}
and $d=4$:
\begin{eqnarray}
\label{bdry4begin}
\phi &=& \mu + \rho u^2 + \frac{w_2^2 \mu}{24} u^6 + O(u^8) \ , \\
w &=& w_2 u^2 - \frac{w_2 \mu^2}{8} u^4 + O(u^6) \ , \\
s &=& 1 - \frac{2 w_2^2 \lambda^2}{9} u^6 + O(u^8) \ , \\
g &=& 1 +g_4 u^4 +  \frac{w_2^2 \lambda^2}{9} u^6 +O(u^8) \ , \\
f &=&1 + f_4 u^4 + \frac{2\lambda^2}{3} ( w_2^2 + \rho^2) u^6 + O(u^8) \ .
\label{bdry4}
\end{eqnarray}
At the horizon, we demand $f(u_h) = 0 = \phi(u_h)$ while the remaining functions $w$, $s$, and $g$ should all be finite. We proceed to solve the differential equations \eqref{eqsuperfluid} by means of a shooting method. If we expand the functions near the horizon we find that there are only four independent coefficients, $\lbrace \phi_1^h , w_0^h , s_0^h , g_0^h \rbrace$, where $\phi_1^h \equiv \phi'(u=u_h)$, $w_0^h \equiv w(u=u_h)$, and similarly for the other two. The method then consists in choosing boundary data $\lbrace \phi_1^h , w_0^h , s_0^h , g_0^h \rbrace$ and (numerically) integrating the differential equations. Once done, we scan the space of solutions in search of the ones with the right boundary values, as in (\ref{bdry3begin}-\ref{bdry3}) and (\ref{bdry4begin}-\ref{bdry4}), in particular obeying $w(u=0)=0$. 
Note that we are picking the boundary metric to be Minkowski
such that $g(u=0) = s(u=0) = 1$. We have used this shooting method to find solutions in $d=3$ and $d=4$ and the goal of this section is to plot the corresponding phase diagrams.\footnote{%
 Ref.\ \cite{Ammon:2009xh} was the first to study the p-wave superconductor with back-reaction in $AdS_5$.  Ref.\ \cite{Gubser:2010dm} provides a corresponding discussion in $AdS_4$.  In \cite{Arias:2012py}, the $AdS_4$ case is also studied. However, the range of parameters where the first order phase transition occurs is not fully explored. Our phase diagrams agree with Fig.~2 of \cite{Gubser:2010dm} and Fig.~8 of \cite{Erdmenger:2012ik}.
 }
  
  Once we have the solutions with the appropriate asymptotic behaviour, the first thing we can plot is the order parameter $ \langle J_x^1 \rangle$ as a function of the temperature. From the AdS/CFT dictionary, we know that
 $\langle J_x^1 \rangle \sim w_1$ and $\langle J_x^1 \rangle \sim w_2$ in $d=3$ and $4$, respectively. The temperature can be read from the periodicity of the time--like direction, and for this phase it reads
\be 
T = \frac{s_0^h}{4 \pi} \left(d - \frac{\lambda^2}{d-1} \frac{(\phi_1^h)^2}{(s_0^h)^2} \right) \ .
\ee
The form of the curve $\langle J_x^1 \rangle$ as a function of the temperature immediately tells us whether we are looking at a first or second order phase transition. 

Below we plot $w_1/\mu^2$ as a function of $T/T_c$ for $\lambda=0.4$, $\lambda=0.8$, and for the critical $\lambda$ where the transition goes from second to first order, which we estimate to be at $\lambda^ {\text{3d}}_c = 0.62 \pm 0.01$. In the dot-dashed green plot we see a solution with a non-vanishing condensate emerges below a certain critical temperature. In the dashed purple one we see that below a certain temperature there are two different solutions with non-vanishing $w_1$, and the transition is first order because the superfluid solution becomes thermodynamically preferred starting at a non--zero value of the order parameter. The results in $d=4$ share the same qualitative profile, as shown in \cite{Ammon:2009xh}, and the critical point can be estimated, $\lambda^ {\text{4d}}_c = 0.365 \pm 0.01$.
\\
\begin{figure}[h] 
\begin{center}
\includegraphics[width=10cm]{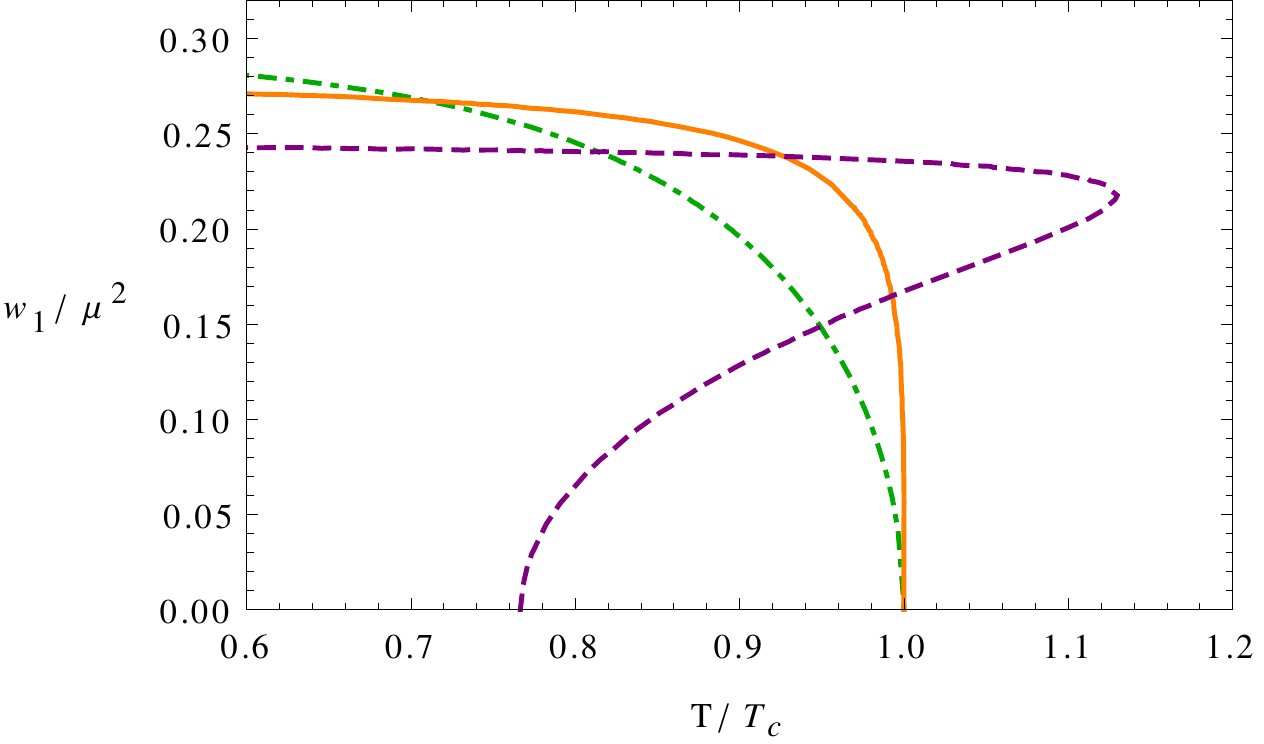}
\end{center}
\caption{The order parameter $\langle J_x^1\rangle \sim w_1$ for different values of $\lambda$, in $d=3$. The curve $\lambda=0.4 < \lambda^{\text{3d}}_c$ (dot-dashed, green) corresponds to a second order phase transition.  
The curve $\lambda=0.8 > \lambda^{\text{3d}}_c$ (dashed, purple) corresponds to a first order phase transition.
The curve $\lambda \sim \lambda^{\text{3d}}_c \sim 0.62$ (solid, orange) passes through the critical point separating the first and second order transitions.
\label{J1ofT}
}
\end{figure}

In order to identify the critical temperatures (which we have already used in Fig.~\ref{J1ofT}) and draw the phase diagram, we need to look at the free energy. The field theory stress tensor is \cite{Balasubramanian:1999re}:
\be
T_{\mu}^{\nu} = \lim_{u\to 0} \frac{1}{ \kappa^2} \sqrt{-\gamma} \left[ - K_\mu^\nu + \left(K + \frac{d-1}{L}\right) \delta_\mu^\nu \right] \ ,
\ee
where $K_{\mu\nu} = \frac{1}{2} (n_{\mu;\nu} + n_{\nu;\mu})$ is the extrinsic trace of a constant $u$ surface, $\gamma_{\mu\nu}$ is the corresponding induced metric, $n^\mu$ is an inward pointing unit vector normal to the surface, and $K = K_\mu^\mu$. We obtain 
\begin{eqnarray}
T_{tt} &=& \frac{L^{d-1}}{2 \kappa^2} (d-1) f_d\ , \\
T_{xx} &=& \frac{L^{d-1}}{2 \kappa^2}(f_d + 2d (d-2) g_d)  \, , \\
T_{y_iy_i} &=& \frac{L^{d-1}}{2 \kappa^2}(f_d - 2d g_d)  \quad i=1, \dots d-2 \,.
\end{eqnarray}

\begin{figure}[h] 
\begin{tabular}{cc} 
\includegraphics[width=7.8cm]{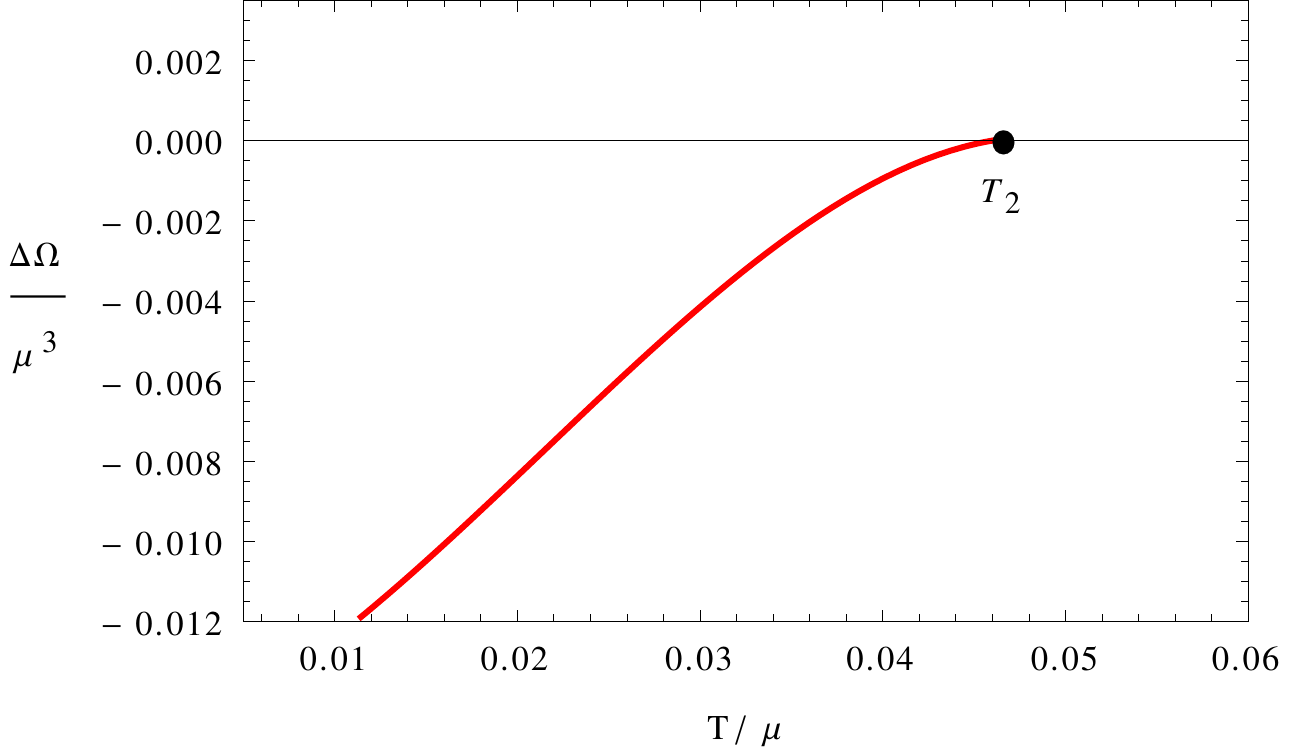} \quad
\includegraphics[width=7.6cm]{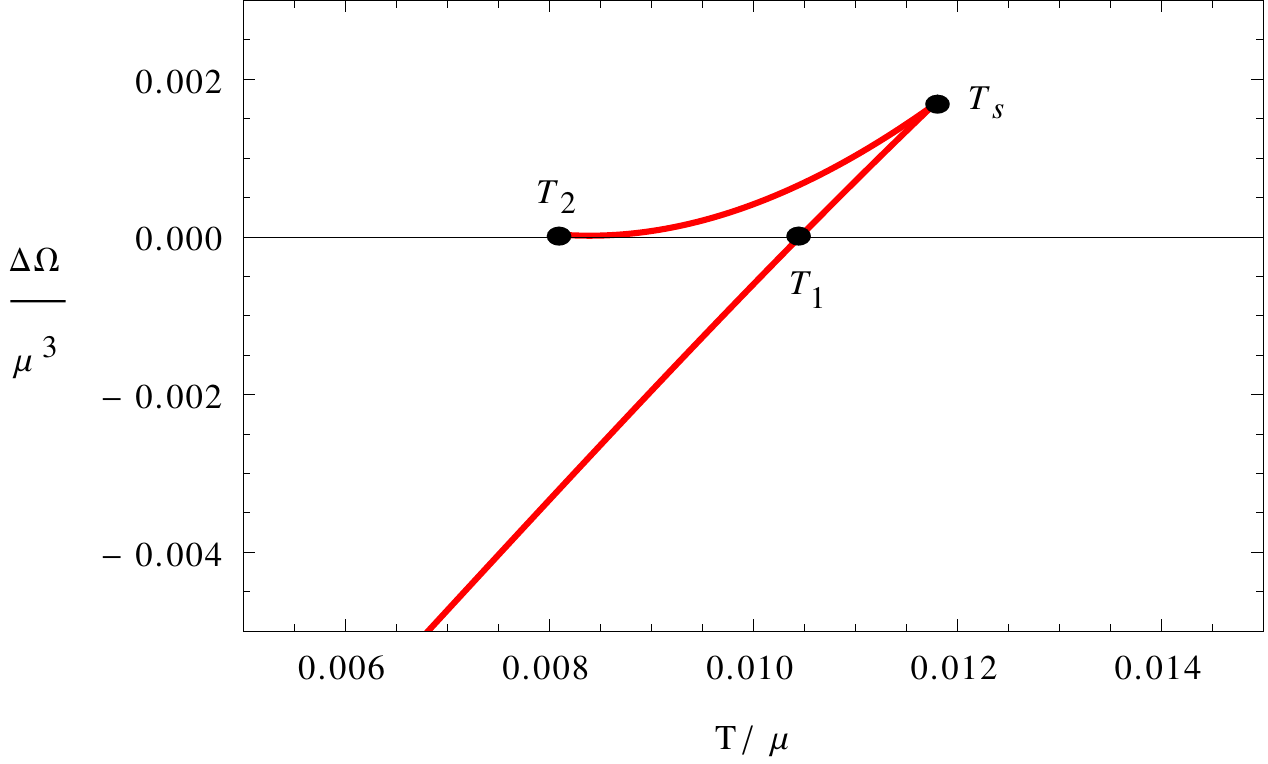}
\end{tabular}
\caption{The difference between the free energies of the superfluid and normal phases in $d=3$, for $\lambda=0.4$ (left) and $\lambda=0.8$ (right). The phase transitions occur at $T_2$ (left), and $T_1$ (right), and are of second and first order, respectively.
On the right, $T_2$ and $T_s$ mark spinodal points in the phase diagram.
\label{DeltaF}
}
\end{figure}

The indices of $T_{\mu\nu}$ are raised and lowered with the Minkowski metric tensor $\eta_{\mu\nu} = (-+\cdots +)$, and with some care the coefficients $f_d$ and $g_d$ can be extracted from the numerical solution near the boundary. To compute the on-shell action $S_{\rm bulk} = \int {\mathcal L}  \, d^{d+1} x$, we note that
\be
{\mathcal L} = -2 \sqrt{-g} \, G^y_y -  \frac{L^{d-1}}{\kappa^2} \left( \frac{fs}{u^{d-2} g }  \left(\frac{g}{u}\right)' \right)' \ .
\ee
We need to add counter-terms to regulate the divergences at $u=0$, 
\be
S = S_{\rm bulk} +  \frac{1}{\kappa^2} \int  \left( K + \frac{(d-1)}{L} \right) \sqrt{-\gamma} \, d^d x  \,,
\ee
where the second term is the Gibbons--Hawking term. We find that in general, $S_{\rm os} = T_{y_iy_i} \operatorname{Vol}/T$ where $\operatorname{Vol}$ is the spatial volume and $T$ is the temperature.  
The free energy is then defined as $\Omega = -T \, S_{\rm os} / \operatorname{Vol}$. An interesting feature, as discussed in \cite{Donos:2013cka} and confirmed by our numerics, is that $g_d=0$. As a consequence, the stress tensor is spatially isotropic, and we can study the object $\widetilde{\Omega} = f_d$ to determine the nature and location of the phase transition. In Fig.\ \ref{DeltaF}, we plot the difference between the free energies of the superfluid and normal phases, $\Delta \widetilde{\Omega}/\mu^3$, as a function of $T/\mu$, for the same values we chose above, $\lambda=0.4$ and $\lambda=0.8$, in $d=3$. Once again the behaviour in $d=4$ is entirely analogous.
%

We can see a clear difference between the first and second order transitions. In Fig.\ \ref{DeltaF} (left), we can identify the second order phase transition temperature $T_2$.  In Fig.\ \ref{DeltaF} (right), in contrast, we can identify instead a first order phase transition temperature $T_1$, as well as two spinodal temperatures $T_s$ and $T_2$.

\begin{figure}[h]
\begin{tabular}{cc}
\includegraphics[width=7.6cm]{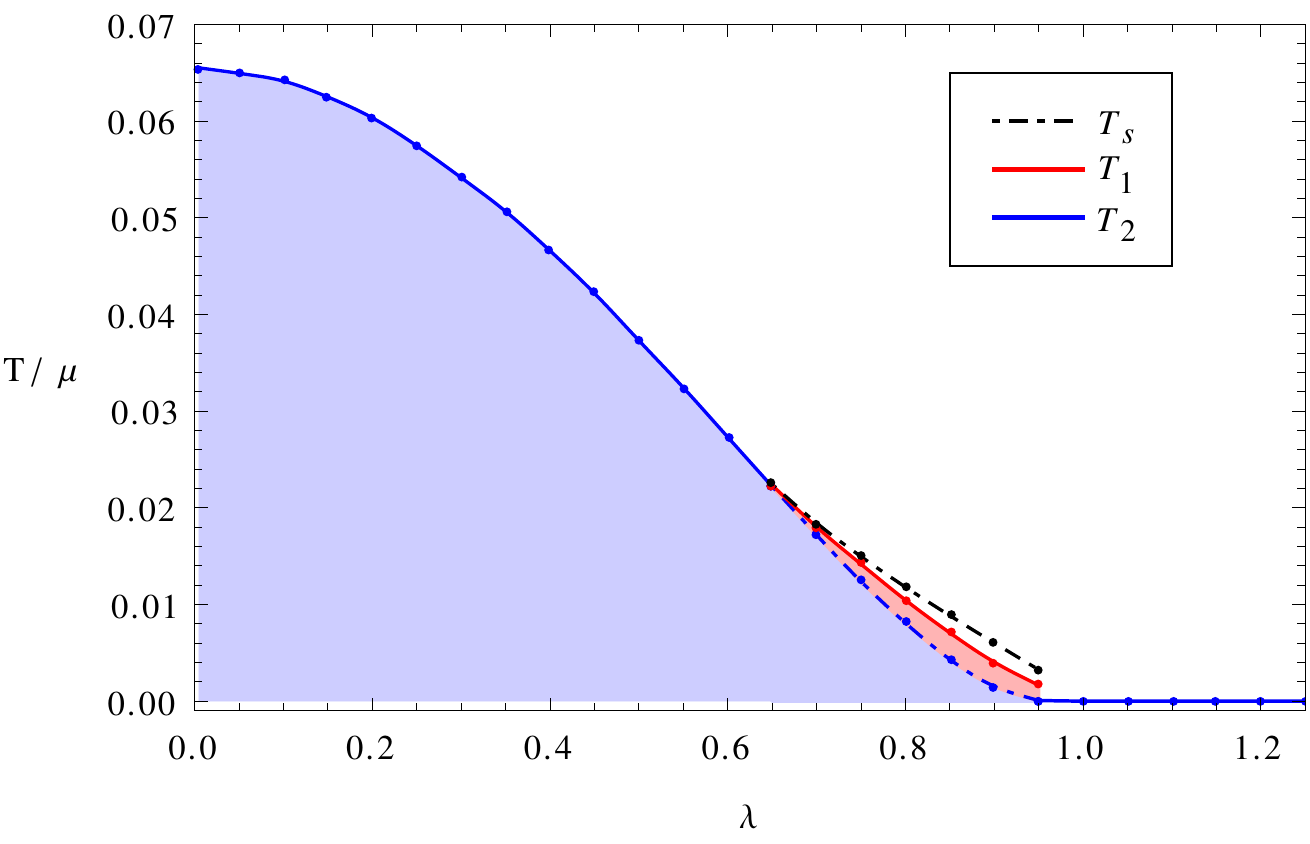} \quad
\includegraphics[width=7.6cm]{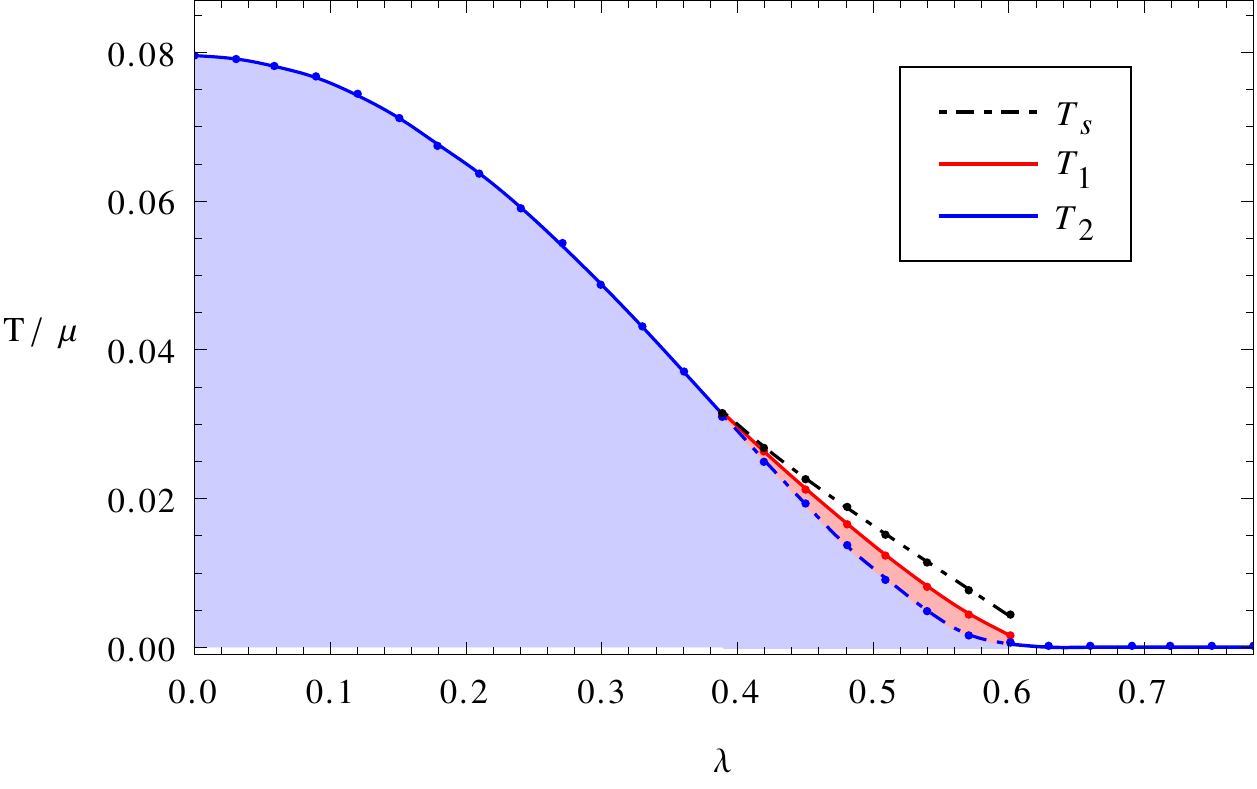}
\end{tabular}
\caption{Phase diagrams of the p--wave superconductor in $(2+1)$ (left) and $(3+1)$ (right) dimensions. For $\lambda<\lambda^{\text{3d}}_c \sim 0.62$ (resp. $\lambda<\lambda^{\text{4d}}_c \sim 0.365$) , there is a second order transition at the solid blue line. For $\lambda > \lambda^{\text{3d,4d}}_c$ the phase transition is first order and occurs at the solid red line. There are two additional spinodal lines (dashed, blue and black), corresponding to temperatures identified in Fig.\ \ref{DeltaF}. The phase diagram on the right coincides with Fig.~$8$ in \cite{Erdmenger:2012ik}.
\label{phasediagram}
}
\end{figure}

By repeating this calculation for different values of $\lambda$, we determined numerically how these special temperatures depend on $\lambda$ and constructed a phase diagram for the holographic p-wave superconductor in the $T/\mu$-$\lambda$ plane.  See Fig.\ \ref{phasediagram} for $d=3$ (left) and $d=4$ (right).
Before the critical point $\lambda_c$, the blue line in Fig.\ \ref{phasediagram} signals a second order phase transition from the normal to the superfluid phase, with a non--vanishing expectation value for the condensate emerging below the critical temperature, as we see in Fig.~\ref{J1ofT} (dot-dashed green). To the right of the critical point, there is a first order transition signalled by the red line. It is of first order because, as we can see in Fig.~\ref{J1ofT} (dashed purple) and \ref{DeltaF} (right), one of the solutions with non--vanishing condensate becomes thermodynamically favoured for a value $\langle J_x^1 \rangle \neq 0$. Beyond the critical point, the blue and black dashed lines represent spinodal lines, with the temperatures clearly identifiable in Fig.~\ref{DeltaF} (right). It bears mentioning that our numerical procedure in the superfluid phase gets harder, and less reliable, once the critical temperature gets too close to zero. 

We note that the blue lines in the phase diagram, both before and after the critical point, can be found in the normal phase through studying the DC conductivity, which we discuss in the next section.

\section{Conductivity for $AdS_4/CFT_3$}

\subsection{Nonabelian Gauge Transformation}
As mentioned earlier, in the normal phase the only nonzero component of the vector potential is $A_t^3 \equiv \phi(u)  =  \mu + \rho \rv^{d-2} $.
We now consider a nonabelian gauge transformation on the background:
\be
A \to \widetilde A = U^{-1} A \, U + U^{-1} \rmd U \ , \label{atransf}
\ee
where
\be
A = -\rmi \, T_a A^a_\mu \, \rmd x^\mu \qquad \mbox{and} \qquad U = \exp \left(\rmi \, T_a \lambda^a\right) \ .
\ee
We here take a particularly simple isospin rotation
\be
\lambda = (0, \theta, 0) \ .
\ee
For this gauge transformation, 
\be
U =
\left(
\begin{array}{cc} 
\cos(\theta/2) & \sin(\theta/2)  \\
-\sin (\theta/2)  & \cos(\theta/2) 
\end{array}
\right) \label{utransf}
\ee
and a short calculation reveals 
\begin{eqnarray}
\left(\widetilde A_t^1,\widetilde A_t^2,\widetilde A_t^3 \right) = \left(A_t^3 \sin \theta, 0, A_t^3\cos \theta \right)  \,.
\end{eqnarray}
This gauge transformation affects the boundary behavior of $A_t$, changing the direction of the chemical potential and charge density vectors in isospin space.  In the context of our holographic application, we assume that the electric field is applied in the 3 direction in the tilde'd coordinate system.  More importantly, there will be an angle $\theta$ in isospin space between the charge density and the electric field.  By tuning $\theta$ to a special value $\theta^*$ we will show it renders a DC conductivity finite, $\sigma_{DC} < \infty$. 

One might na\"{i}vely guess this goal is achieved when $\theta^* = \pi/2$, when the electric field and charge density are orthogonal in isospin space. However, because of the non-abelian terms in the action, $\theta^*$ will depend on temperature, ranging from $\theta^* = 0$ at the second order phase transition (blue lines in the phase diagrams in Fig.\ \ref{phasediagram}) all the way to a finite value that interpolates between $0$ (at $\lambda=0$) and $\pi/2$ (as $\lambda \to \infty$) in the high temperature limit. We will explore the special $\theta^*$ in more detail in the following section for $d=3$. 

\subsection{Normal Phase Conductivity} \label{normal}

We consider fluctuations around the normal phase background, $a_x^a$ and $g_{tx}$, and assume they are small.  
We work in the untilde'd frame in order to keep the background solution as simple as possible, and transform to the tilde'd frame at the end.  
While we let the fluctuations have arbitrary radial dependence, we restrict the time dependence to have the form $\rme^{-\rmi \omega t}$.  At linear order, the equation of motion for the metric fluctuation $g_{tx}$ is 
\be
\frac{1}{\rv^2} \partial_\rv (\rv^2 g_{tx}) =-2 \lambda^2 \phi' L^2  a_x^3 \ . \label{ads4eom1}
\ee
The equations of motion for the gauge fields are then
\begin{eqnarray}
\label{fluctnormal1}
u^{d-3} f \, \partial_\rv \left( f u^{3-d} \partial_\rv a^3_x \right) &=& \left( -\omega^2 + 2 \lambda^2 (\phi')^2 u^2 f \right) a_x^3 \ , \\
u^{d-3} f \, \partial_\rv \left( f u^{3-d} \partial_\rv a_\pm \right) &=& -\left( \pm \omega  -\phi \right)^2 a_\pm \ .
\label{fluctnormal2}
\end{eqnarray}
where we have defined $a_\pm \equiv a_x^1 \pm \rmi a_x^2$.

The way in which a finite DC conductivity can be extracted from equations (\ref{fluctnormal1}) and (\ref{fluctnormal2}) can be understood at a schematic level.  Given an electric field in the 3 isospin direction, the pole in the DC conductivity at $\omega=0$ comes from the $(\phi')^2 a_x^3$ term in (\ref{fluctnormal1}).  Similarly, for an electric field in the 1 or 2 isospin directions, the pole (this time with the opposite sign) would come from the 
$-\phi^2 a_\pm$ term.  By carefully selecting the angle $\theta$, we can cancel out one pole with the other.
It might be somewhat counter-intuitive that the DC conductivity remains finite in the absence of momentum dissipation, but we should note that while the cancellation of poles leads to a finite DC conductivity in the isospin direction parallel to the electric field, $\vec E$ will act to accelerate a current in an orthogonal direction in isospin space\footnote{%
  More specifically, the chemical potential and $\vec E$ pick out a plane in isospin space.  $\vec E$ will act to accelerate a current in this plane but orthogonal to $\vec E$.
}
 and also in energy density. 

At a physical level, the $(\phi')^2 a_x^3$ term in (\ref{fluctnormal1}) comes from the mixing of the momentum and charge currents.  Indeed, using eq.\ (\ref{ads4eom1}), $\phi'$ can be replaced with $g_{tx}$.  The $-\phi^2 a_\pm$ term in (\ref{fluctnormal2}) appears because $a_\pm$ acts like a charged particle  under $F_{\mu\nu}^3$.  The time derivative is shifted by the connection term $a_t^3 = \phi$.  A similar mass-like term $a_x \psi^2$ produces a pole in the DC conductivity for the s-wave holographic superconductor \cite{Hartnoll:2008vx}.  Here $\psi$ is the scalar field which develops a nontrivial profile below the superconducting phase transition temperature.

We now compute the conductivity by solving equations \eqref{fluctnormal1} and \eqref{fluctnormal2} numerically.  Focusing on $d=3$, we integrate from the horizon, setting  $u_h=1$, to the boundary, imposing the following near--horizon expansions 
\be 
a_x^i = (1-u)^{-\frac{\rmi \omega}{4 \pi T}} \left(a_0^i + a_1^i \, (1-u) + a_2^i (1-u^2) + \dots \right) \,; \,\, i=1, 2, 3 \,.
\label{nearhorizon}
\ee
We here consider generalized isospin-rotated backgrounds such that the tilde'd gauge field components are related to untilde'd gauge fields, using \eqref{atransf} and \eqref{utransf}, in the following way:
\begin{eqnarray}
\label{tildeda}
\widetilde{a}_x^1 &=& \cos\theta \,a_x^1+ \sin\theta \,a_x^3\ , \\
\widetilde{a}_x^2 &=& a_x^2\ ,  \\
\widetilde{a}_x^3 &=& -\sin\theta \, a_x^1+\cos\theta \, a_x^3 \ , \label{ax3transf}
\end{eqnarray}
where $\theta$ is the rotation angle discussed above. Since we consider that experiments are measured in the tilde'd frame, we first require our boundary conditions to be $\widetilde{a}_x^1 \big|_{u=0}=\widetilde{a}_x^2\big|_{u=0}=0$, so that there is only an electric field along the third isospin direction, 
${\cal E} = \rmi \, \omega \widetilde{a}_x^3 \big|_{u=0}$.
We will be interested in computing the conductivity defined as 
\begin{eqnarray}
\label{sigma}
\sigma(\omega,\theta) = \frac{1}{\rmi \, \omega} \frac{\widetilde{a}^{3 \,(1)}_{x}}{\widetilde{a}^{3 \,(0)}_{x}} \equiv \frac{\cal J(\omega ,\theta) }   {\cal E(\omega ,\theta)}  \ ,
\end{eqnarray}
where $\cal{J}$ and $\cal {E}$ are current and applied electric field, respectively, and
the gauge fields have near-boundary expansions of the form $a_{\mu}^i(u) = \left(a_{\mu}^i\right)^{(0)} + u^{d-2} \left(a_{\mu}^i\right)^{(1)} + \ldots$. 

The way to obtain a finite DC resistivity is as follows.  
We fix the additional freedom from the isospin rotation angle $\theta$ by requiring ${\cal J}(\omega=0 ,\theta^*)=0$ for a special angle $\theta^*$. Schematically we can write the derivatives 
\be 
\left(a_{x}^1\right)^{(1)} = q \sin \theta \,\,;  \quad \left(a_{x}^3 \right)^{(1)} = p \cos \theta \,. \label{normal_der}
\ee
Going back to \eqref{ax3transf}, we see that the condition ${\cal J}(\omega=0 ,\theta^*)=0$ means that $\theta^*$ is a solution to the equation 
\be 
\cos^2 \theta =\left(1 + \frac{p}{q} \right)^{-1} \Big|_{\omega=0} \,. \label{theta_normal}
\ee
This choice in turn will imply that $\sigma_{\text{DC}} = \lim_{\omega \rightarrow 0} \sigma(\omega)$ is finite, as the first term in ${\cal J}(\omega ,\theta^*)$ becomes of ${\cal O}(\omega)$, cancelling the $\omega$ in the denominator. An interesting feature is that this parameter $\theta^*$ can help us locate the phase boundary. 
The existence of a zero mode for $a_x^1$ along the blue curve (solid and dashed) in the phase diagram in Fig.\ \ref{phasediagram} implies that the condition ${\cal J}(\omega=0 ,\theta^*)=0$
can be satisfied by a smaller and smaller $\theta^*$ as one approaches the phase boundary from the normal phase. 
As one decreases the temperature through the phase boundary, $1/q$ passes through zero from above while $p$ remains positive and order one.
If we then go beyond the phase boundary but insist on remaining in the now unstable normal phase, numerically we find
we cannot
satisfy ${\cal J}(\omega=0 ,\theta^*)=0$, since we get a condition $\cos^2 \theta  > 1$ which cannot be satisfied by a real $\theta^*$. 

We are able to give an analytic form of the special $\theta^*$ in the limit of $T/\mu \to \infty$. 
We solve the equations of motion \eqref{fluctnormal1}, \eqref{fluctnormal2} in powers of $\mu$, order by order.  The expansion is  
\be 
a_x^i(\rv) = a_x^{i,0}(\rv) + \mu^2 a_x^{i,1}(\rv) + \dots \ .
\ee
At order $\mu^0$, taking into account the boundary conditions implied by the gauge transformation and imposing regularity at the horizon, we get constant solutions 
\be
a_x^{1,0}(\rv) = -{\cal C} \sin\theta \,, \quad a_x^{3,0}(\rv) = {\cal C} \cos\theta \,.
\ee
At order $\mu^2$, we can directly integrate the differential equations (\ref{fluctnormal1}) and (\ref{fluctnormal2}) to get the corrections:
\begin{eqnarray}
a_x^{3,1} &=& 2 \lambda^2 {\cal C} \cos \theta \int_0^u \frac{1}{f(u_1)} \int_1^{u_1}  u_2^2  du_2 du_1 \ , \\
a_x^{1,1} &=& - {\cal C} \sin \theta \int_0^u \frac{1}{f(u_1)} \int_1^{u_1} \frac{(1-u_2)^2}{f(u_2)} du_2 du_1 \ .
\end{eqnarray}

\begin{figure}[h]
\begin{center}
\includegraphics[width=10cm]{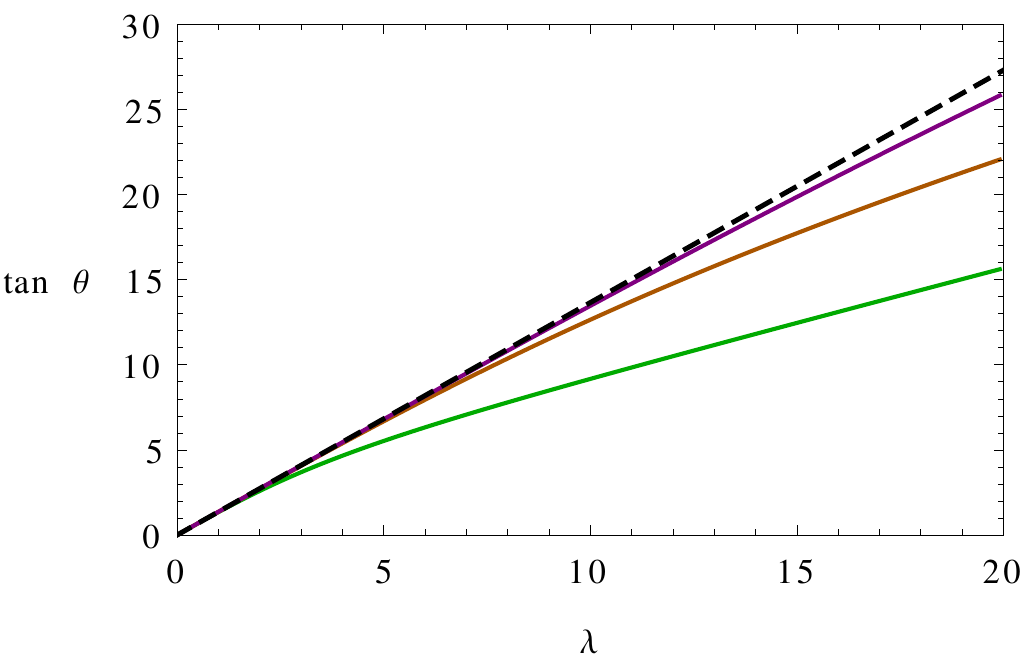} 
\end{center}
\caption{$\tan \theta$ for the special angle $\theta^*$ that makes $\sigma_{DC}$ finite, as a function of $\lambda$, for $d=3$. The three solid lines, from bottom to top, correspond to fixed $T/\mu = 1, 4, 10$.
The dashed line, $\tan \theta^* = a_3 \, \lambda$, where $a_3 = \frac{4}{\sqrt{3} \,\pi + 3 \log(3)}$ is the high temperature limit $T/\mu \to \infty$. 
\label{theta_inf}
}
\end{figure}

With these results in hand, we can read off what we defined as $p$ and $q$ above (at leading order in powers of $\mu$)
\be 
p/{\cal C} = -\frac{2}{3} \lambda^2 \,; \quad q/{\cal C} = \frac{1}{2} \log(3) - \frac{\pi}{2 \sqrt{3}} \,.
\ee
In the limit $T/\mu \to \infty$ the special parameter $\theta^*$ can be described by the following expresion:
\bea
\tan \theta^* = a_3 \lambda , 
\eea
with
\bea
a_3 = \frac{4}{\sqrt{3} \,\pi + 3 \log(3)}  .
\eea 
We have confirmed this result numerically in Fig.~\ref{theta_inf}. We see that as $T/\mu$ increases the curves move toward the straight line with slope $a_3$.

\begin{figure}[h]
\begin{tabular}{cc} 
\includegraphics[width=8cm]{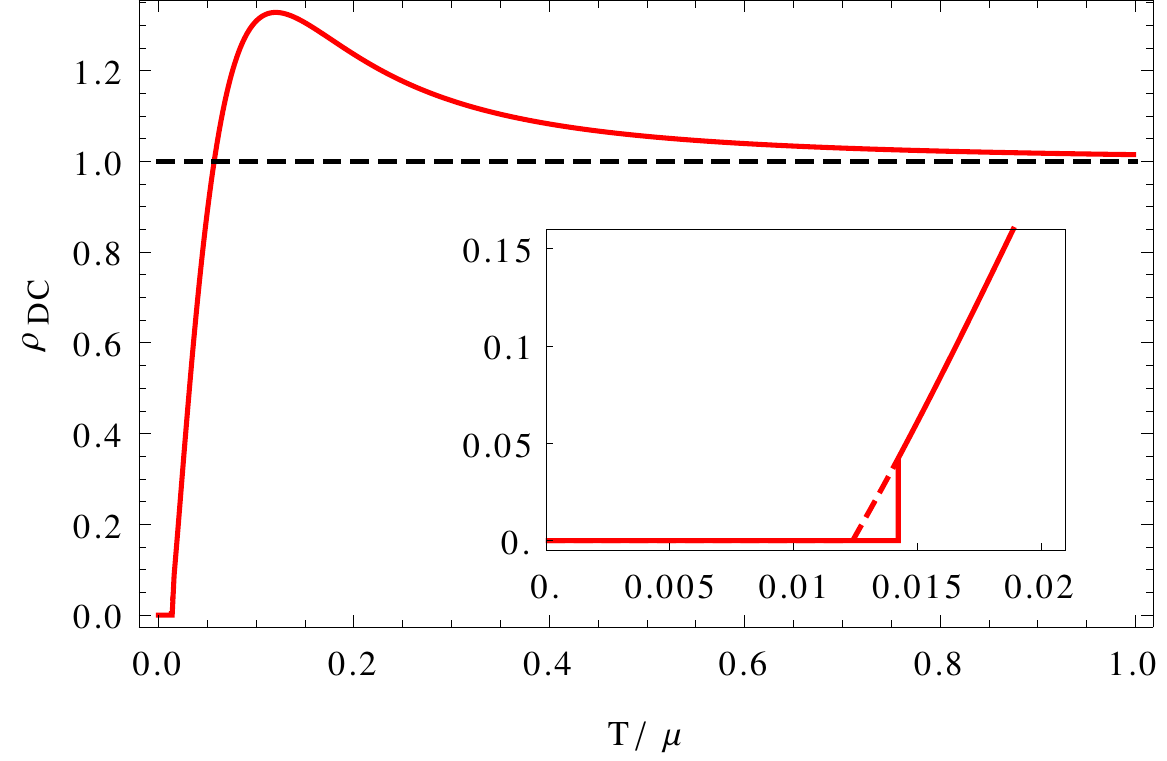}
\includegraphics[width=8cm]{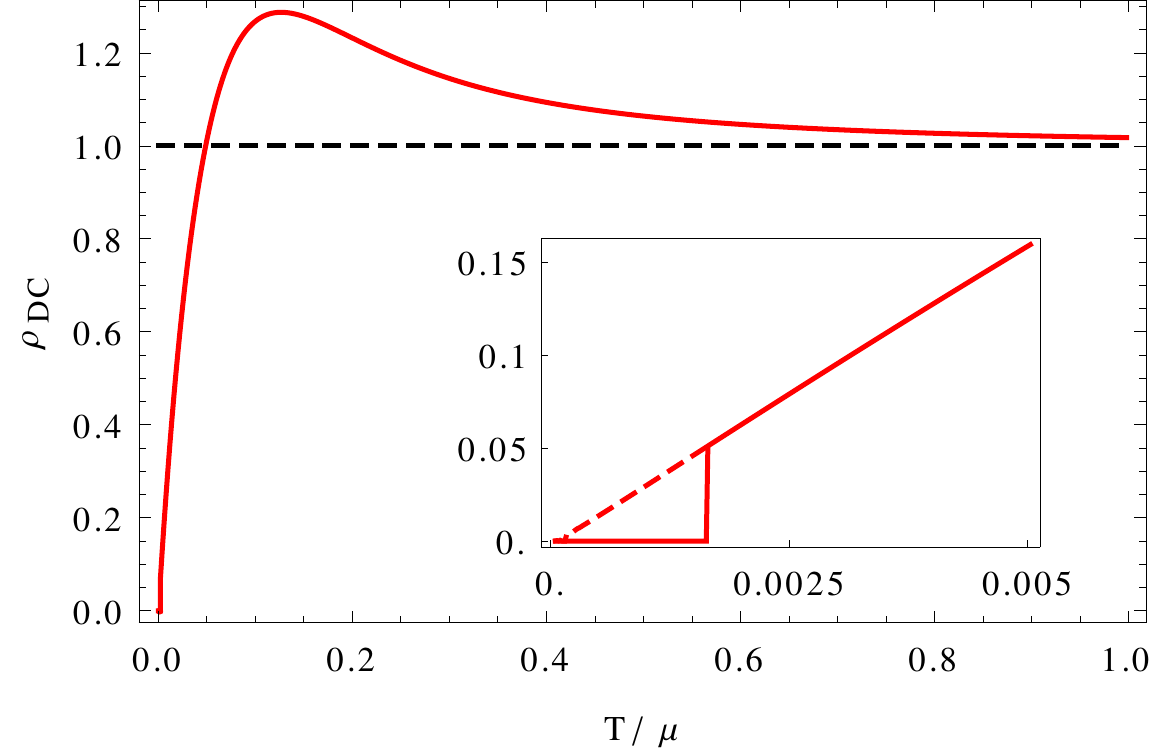}
\end{tabular}
\caption{DC resistivity versus $T\over \mu$, for $\lambda=0.75$ (left) and $\lambda=0.95$ (right), in $d=3$. In the main plots we show the resistivity over a wide range of temperatures.
In the insets, we show what happens close to ${T \over \mu} =0$
\label{rhoofT}
}
\end{figure}

We next plot the DC resistivity $\rho_{\text{DC}}$ as a function of the temperature. Notice that for $\lambda > \lambda_c$, the transition to the superconductor solution becomes of first order.  The consequence of the first order transition is that $\rho_{\text{DC}}$ should exhibit a step, going abruptly from a finite value to zero.\footnote{%
We have implicitly assumed that $\rho_{DC} = 0$ in the superconducting phase.  In light of our non-abelian gauge transformation, we will revisit this assumption in the next subsection.
}
The small $T$ behavior of the resistivity can be seen in the Fig.\ \ref{rhoofT} insets for $\lambda=0.75$ and $0.95$.
We see that the resistivity jumps from $0$ to a finite number, and that its behaviour is linear right after the phase transition.  Notice that changing $\lambda$ will not affect the linear dependence of the DC resistivity. It only changes the intercept and in particular, when $\lambda\sim 1$ (numerically, we here take $\lambda=0.95$), the resistivity line passes through the point where the second order phase transition temperature is zero. The value of $\lambda$ where $T_2=0$ is the critical point of an underlying quantum phase transition.  Note that our rotation angle $\theta$ is fixed by having a finite DC conductivity. Therefore, with no more free parameters left to fine-tune our plots the linear resistivity feature is robust in our model.
\\

In Fig.\ \ref{rhoofT}, we also show the DC resistivity as a function of $T/\mu$ for larger values of $T$.
We find that while the resistivity is growing, it is well approximated by a profile of the form $\rho_{\text{DC}} = a \frac{T}{\mu} + b \left( \frac{T}{\mu}\right)^{3/2}$. Next the resistivity reaches a maximum before decaying towards $\rho=1$ for larger temperatures.
(At large $T/\mu$, the underlying scale invariance of the 2+1 dimensional 
field theory and dimensional analysis forces $\rho$ to be constant.)\footnote{%
The approach to $\rho = 1$ is found to be well approximated by a function of the form   $a' \left( \frac{T}{\mu}\right)^{-1} + b' \left( \frac{T}{\mu}\right)^{-2}$. 
}

Next we plot the optical conductivity $\sigma(\omega)$ in Fig.\ \ref{ac}. Our optical conductivity does not have a standard Drude peak in the real part of the conductivity. The absence of the Drude peak might be an expected result if we recall that we do not introduce any lattice or impurities in the system; hence the momentum does not dissipate as it does in the Drude model. The real part of the optical conductivity instead shows a drop in the DC limit. It would be interesting to understand this behavior better.  We notice that in the holographic models with a lattice studied recently in \cite{Horowitz:2012ky, Horowitz:2012gs, Horowitz:2013jaa}, the Drude peak is observed and a robust frequency scaling in the ``Drude tail'' region of the optical conductivity is found to agree with some cuprate experiments.  Their resistivity is sensitively dependent on the lattice parameter. 
It would be interesting to introduce momentum relaxation in our model to see how it modifies the low frequency behavior of the optical conductivity and also to see if the linear temperature resistivity can be preserved. We leave this problem as a future project. However, our results suggest that a  linear DC resistivity can be produced in the absence of a lattice. 

\begin{figure}[h]
\begin{tabular}{cc}  \label{ac}
\includegraphics[width=7.6cm]{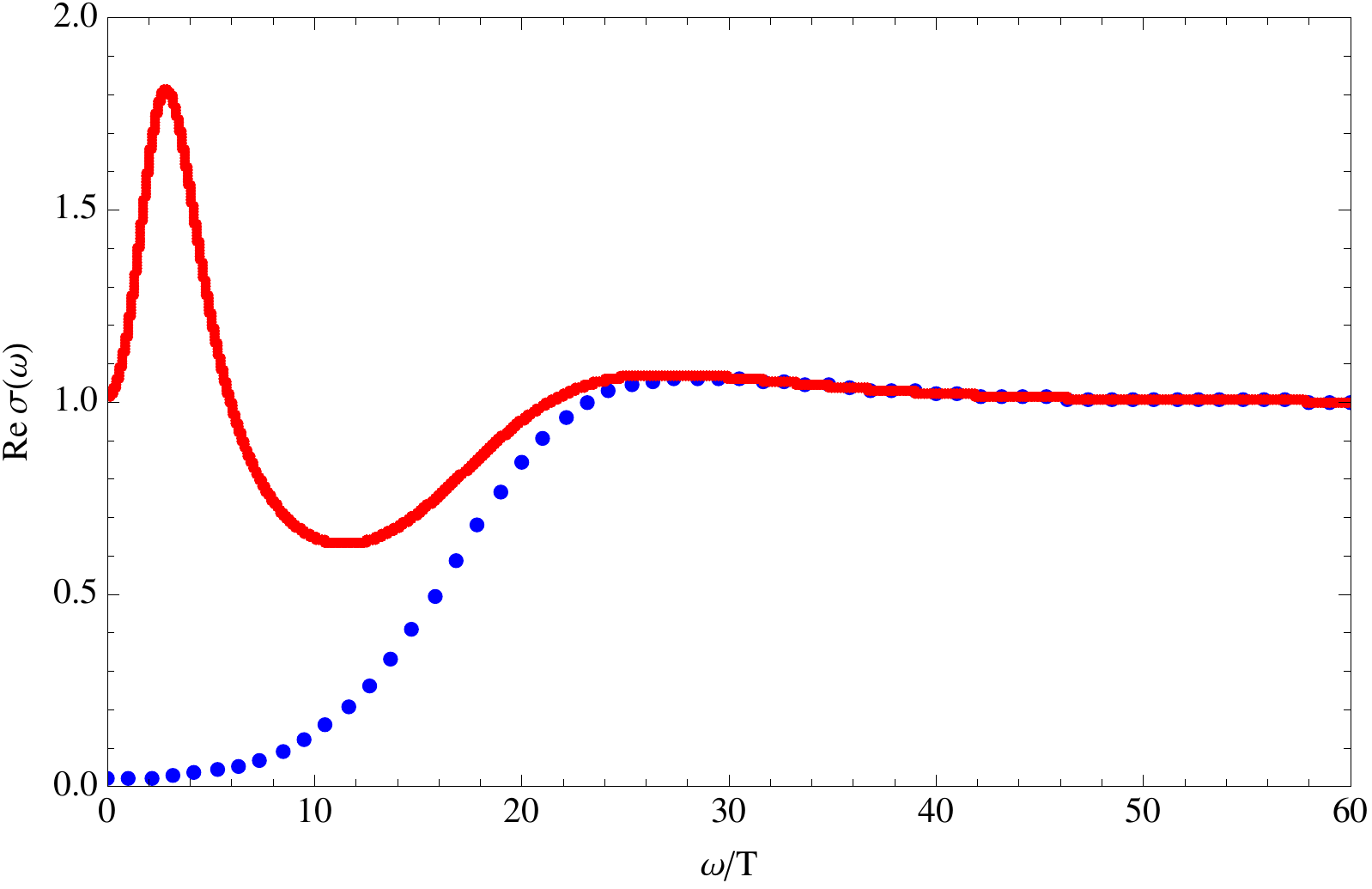} \label{ac}
\includegraphics[width=7.6cm]{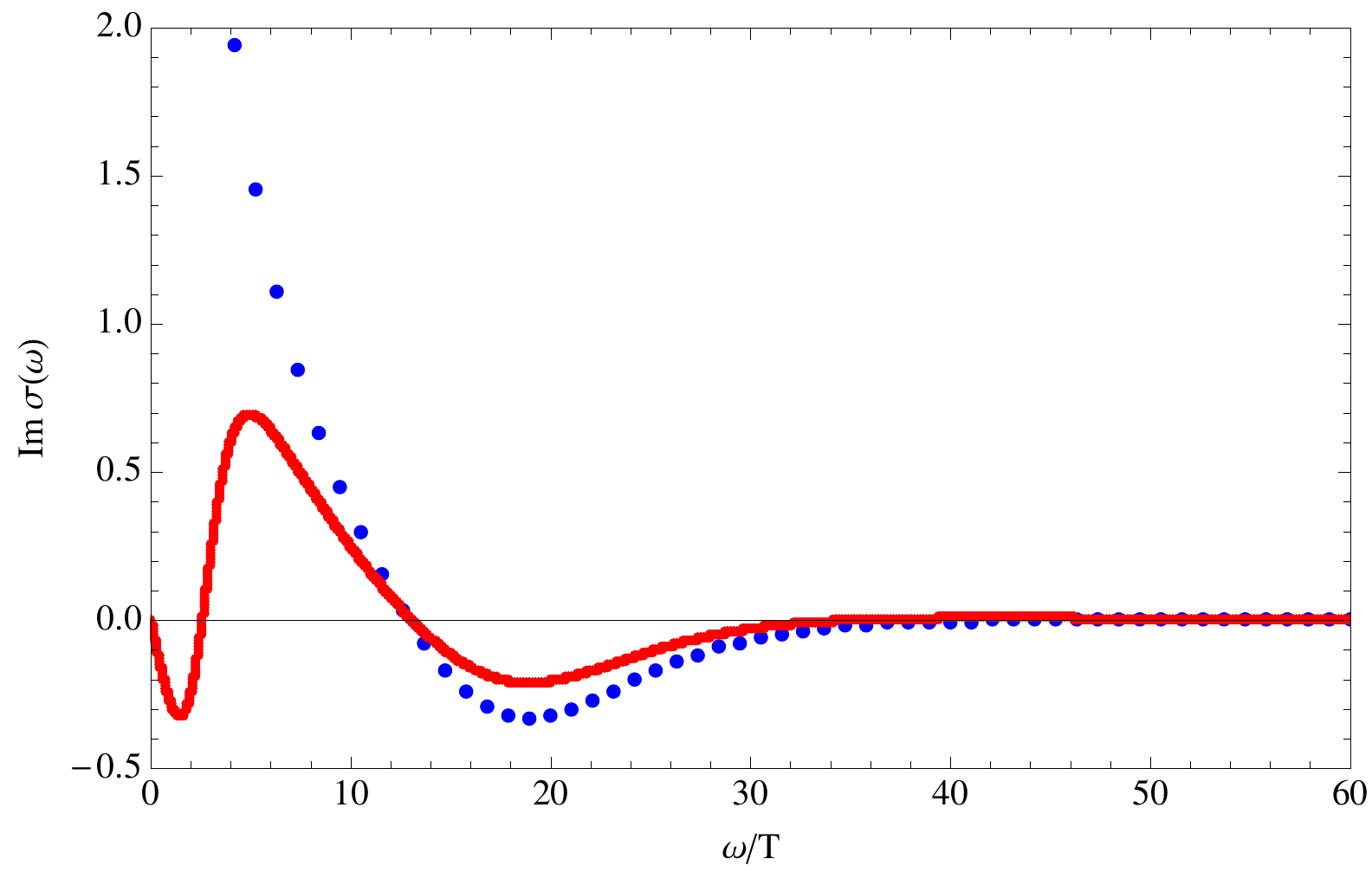}
\label{ac}
\end{tabular}
\caption{The red (solid) lines are the optical conductivity with the special $\theta^*$ that makes the DC conductivity finite. The blue (dotted) lines correspond to $\theta=0$. These plots were made with $\lambda=0.95$, where the second order transition temperature is close to zero, and ${T\over \mu}=0.047$, which is in the normal phase. Plots of different temperature have similar behaviours:  Our optical conductivity does not have a standard Drude peak. 
\label{ac}
}
\end{figure}

\subsection{Superfluid Phase}

A natural question to ask is whether the ``trick'' of finding a special gauge transformation to make the DC conductivity finite can also be done in the superfluid phase. In this section we argue that, close to $T=T_c$ (and where the phase transition is second order), the answer is negative. While we na\"ively expect this property to then hold in the entire superfluid phase, we do not have a proof.

To address this question, we have to study fluctuations around the superfluid background discussed above. This was thoroughly done in \cite{Erdmenger:2012zu}, in the context of $AdS_5$. With the appropriate changes, the entire concept and procedure applies to our solution in $AdS_4$. We introduce the fluctuations $h_{\mu \nu}$ and $a_{\mu}^i$, such that the total metric and gauge field are
\be 
\hat{g}_{\mu  \nu} = g_{\mu  \nu} + h_{\mu  \nu} \,\,; \quad 
\hat{A}_{\mu}^i = A^{i}_{\mu} + a_{\mu}^i \,,
\ee
and the first terms are the background solution.\footnote{Recall that the background had gauge field components $A_x^1 = w(u)$ and $A_t^3 = \phi(u)$. And for simplicity we will set $u_h = L = 1$ from the start.} The fluctuations have space-time dependences of the form $a_{\mu}^i = \rme^ {-\rmi \, \omega \, t} \widehat{a}_{\mu}^i(u)$, but we will drop the hats to simplify the notation. One important feature is that the entire set of fluctuations splits into two blocks. From a practical standpoint this means that the linearised equations of motion decouple into two sets. The set that is of most interest to us contains the fields $\lbrace a_t^1 \,, a_t^2 \,, a_x^3 \,, \xi_{tx} \rbrace$, where we have defined $\xi_{t x} \equiv g^{x x} \, h_{t x} $. We follow the steps of \cite{Erdmenger:2012zu}, writing down the on--shell action involving terms quadratic in these fields, then regularizing with the appropriate counterterms. The situation in our case is less involved than in $AdS_5$, as only a simple counterterm of the form $S_{\text{ct}} = c \int \rmd^3 x \, \sqrt{-\gamma}$, with $c$ a constant and $\gamma$ the induced metric on the boundary, is required to regularize the on--shell action. We then proceed to compute the matrix of Green's functions, from which we can extract the relevant component
\be 
G^{x,x}_{3,3} = \frac{\lambda^2}{\left(a_x^3\right)^{(0)}} \left[ \left(a_x^3\right)^{(1)} + w_1 \frac{\mu \, \left(a_t^1\right)^{(0)} + \rmi \, \omega \, \left(a_t^2\right)^{(0)}}{\mu^2 - \omega^2} \right]\,. \label{Gsuperf}
\ee
The terms in the expression above correspond to terms in the near-boundary expansion of the gauge fields, which takes the form $a_{\mu}^i = (a_{\mu}^i)^{(0)} + u \, (a_{\mu}^i)^{(1)} + \ldots$. The conductivity is related to the result above as
\be
\sigma^{x,x} = \frac{1}{\rmi \, \omega} G^{x,x}_{3,3} 
\ee
We note that the answer looks very similar to the one in \cite{Erdmenger:2012zu} (without the first term), and agrees with the result found in \cite{Gubser:2008wv} working in $AdS_4$ but in the probe limit. The calculation and analysis of $\sigma^{x,x}$, and other transport coefficients, is an involved task from a numerical point of view, as evidenced in \cite{Erdmenger:2012zu}, and is left for future work. However, the possibility of finding a special gauge transformation to make $\sigma^{x,x}_{\text{DC}}$ finite can be addressed independently, and we argue that the answer is negative at least close to $T_c$. 
Like in the normal phase, we have two systems with gauge fields $A^{i}_{\mu} + a_{\mu}^i$, $\widetilde{A}^{i}_{\mu} + \widetilde{a}_{\mu}^i$ related by a gauge transformation. We then imagine that we measure the conductivity in the tilde'd system, and again that there is only an electric field along the third isospin direction. As before, the electric fields in the one and two isospin directions must vanish,
\be 
 \widetilde{a}_{x}^1 (u=0) = \widetilde{a}_{x}^2 (u=0) = 0 \,.
\ee
Setting $\left(\widetilde{a}_{x}^3\right)^{(0)} \equiv a_3$, we can then take 
\be
\left(a_{x}^1\right)^{(0)} = - a_3 \sin \theta \,\,; \quad \left(a_{x}^2\right)^{(0)} = 0 \,\,; \quad \left(a_{x}^3\right)^{(0)} = a_3 \cos \theta \,\,; 
\ee
We also define $\left(\widetilde{a}_{t}^1\right)^{(0)} \equiv  a_1$, with $a_1$ and $a_3$ numbers that are, in principle, related by the differential equations. The finiteness of the DC conductivity corresponds to the Green's function \eqref{Gsuperf} vanishing at $\omega=0$, as we discussed in the normal phase context above. Using the notation introduced in \eqref{normal_der}, this condition is spelled out as
\be 
\left(\left(\widetilde{a}_{x}^3\right)^{(1)} + \frac{\widetilde{w}_1}{\widetilde{\mu}} \left(\widetilde{a}_{t}^1\right)^{(0)}\right)_{\omega=0} = \left(- p \, \cos^2 \theta + q \sin^2 \theta + \frac{w_1}{\mu} a_1 \right)_{\omega=0} = 0 \,.
\ee
Dropping the subscript but remembering that all objects are evaluated at $\omega=0$, we get a condition for $\theta$ that is the generalization of \eqref{theta_normal}
\be 
\cos^2 \theta = \frac{q + \frac{w_1}{\mu} a_1}{p + q} \,. \label{theta_super}
\ee
As we discussed above, in the normal phase ($w_1 = 0$), the phase boundary is hit when $p/q$ crosses $0$, and later becomes negative, so that the \eqref{theta_normal} no longer has a real solution. We now wonder if we can understand what happens once we cross to the superfluid phase, and whether the equation above can be satisfied or not. Close to the phase transition, and for $\lambda < \lambda_c$, $w_1$ is very small, so we can take the superfluid solution to be the normal solution at first approximation, with corrections due to the non--vanishing $w_1$ on top of that. Similarly, we assume that at first approximation the fluctuations are also close to the fluctuations we studied in the normal phase. In that spirit, and with $p/q$ also small (and negative) close to the transition, we approximate \eqref{theta_super} as 
\be 
\cos^2 \theta = \frac{1 - \frac{w_1}{\mu} \frac{a_1}{q}}{1 + \frac{p}{q}} \sim 1 + \left( \left|\frac{p}{q}\right| - \frac{w_1}{\mu} \frac{a_1}{q} \right) + \dots \label{final_theta}
\ee
We now analyse each term separately. The factor $1/q$ crosses $0$ at the phase transition, but more specifically, $q$ behaves as $\left(T_c - T \right)^{-1}$, a feature that we find numerically. On the other hand, $w_1$ behaves as $\left(T_c - T \right)^{1/2}$, as shown in Fig.~\ref{J1ofT}. Finally, knowing that $a_{t}^1$ vanishes when $w_1=0$, we know that the first correction will be proportional to $w_1^k$, with $k>0$. Going back to \eqref{final_theta} with this information, we see that the first term in the bracket is positive and of order $w_1^2$, whereas the second is at least of order $w_1^3$. This means that the right--hand side of \eqref{final_theta} is greater than 1, implying there is no real solution to the equation, and thus no way to make the DC conductivity finite. 

We stress again that this argument is only valid when the phase transition is of second order, since in the first order case we cannot make this approximation with a small $w_1$. Also, as we move away from the second order transition, the expectation value of the condensate can become of ${\cal O}(1)$, so further corrections to the formulae above are no longer suppressed and $w_1$ is no longer a good perturbative parameter. However, we expect this property to be uniform to the entire superfluid phase, so this feature will hold everywhere below the phase transition, even though a full-fledged solution for the fluctuations may be required to prove this statement.

\section{Analytic conductivity for $AdS_5/CFT_4$}

\begin{figure}[h]
\begin{tabular}{cc} 
\includegraphics[width=8cm]{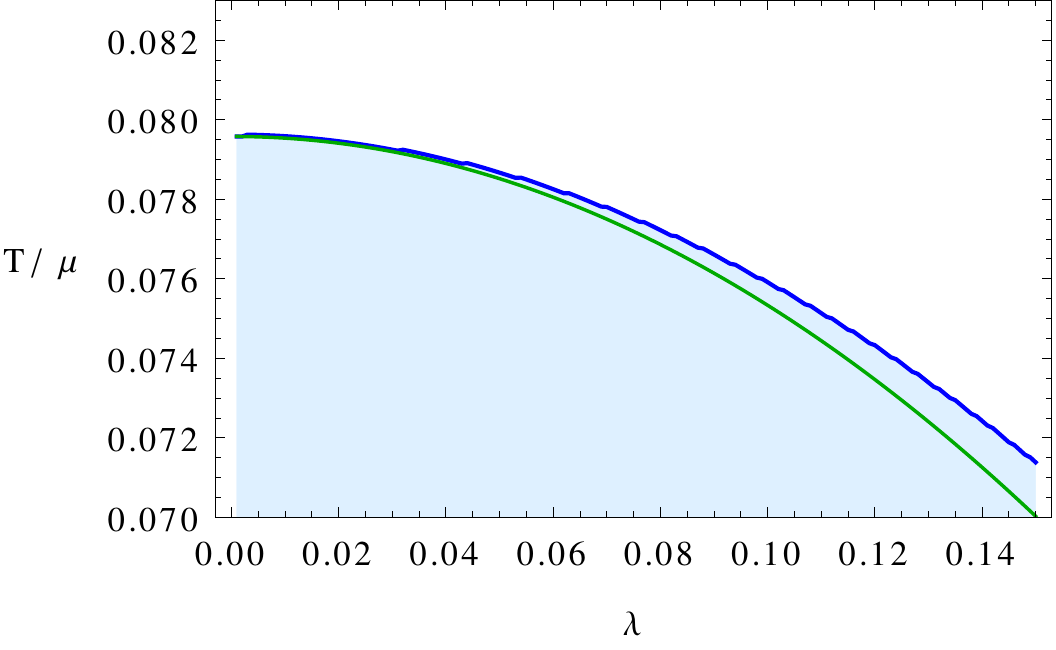} \quad
\includegraphics[width=7.6cm]{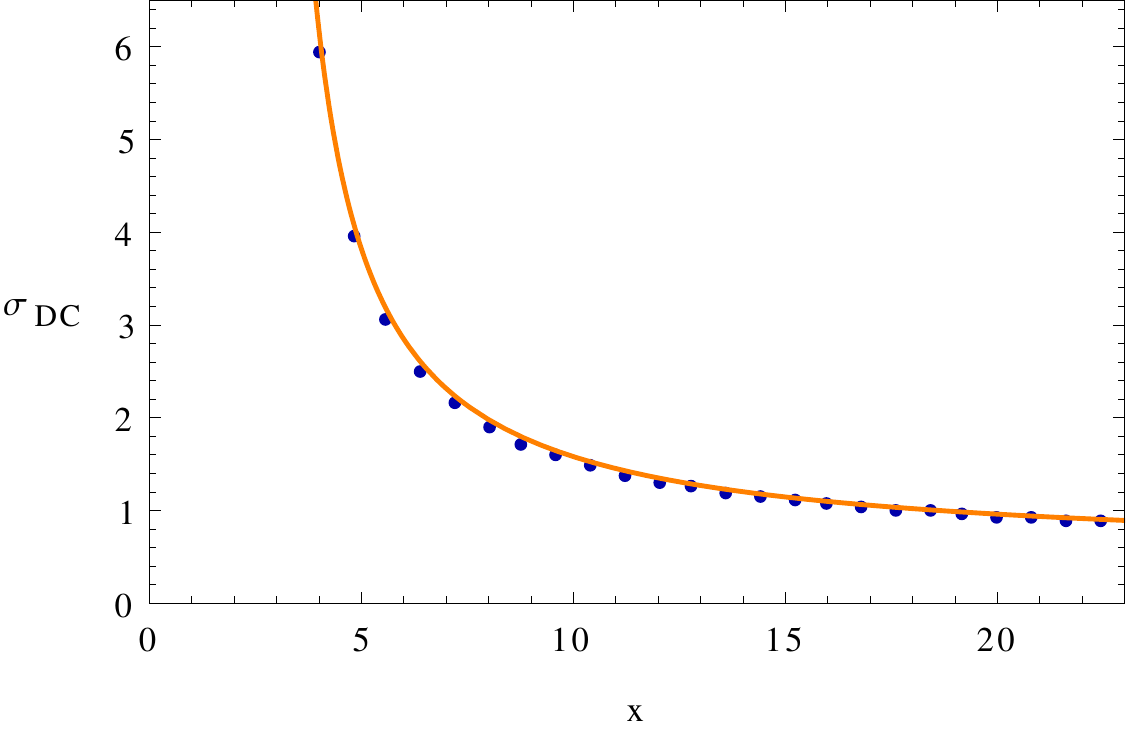}
\end{tabular}
\caption{Left: Phase boundary (blue), and curve $\mu=4$ (green), showing that only one order of limits is allowed in \eqref{ads5}. Right: DC-conductivity versus $x \equiv - \frac{\delta\mu}{\lambda^2}$ with fixed $\delta\mu = -0.01$. The orange line is the analytic prediction \eqref{ads5}, and the blue dots were computed numerically.
\label{mueq4}
 }
\end{figure}

In this section, we give an analytic expression of DC conductivity for $AdS_5/CFT_4$.  
We set $d=4$ in the equations of motion \eqref{ads4eom1}, \eqref{fluctnormal1}, and \eqref{fluctnormal2}.
Setting $u_h = 1$ as before, the horizon series expansion is given by \eqref{nearhorizon}.
Now the nice thing about (\ref{fluctnormal2}) is that an analytic solution is known 
\cite{Basu:2008bh}
when $d=4$, $\mu=4$, $\lambda=0$ and $\omega=0$:
\be
a_\pm = c_1 \frac{u^2}{(1+u^2)^2} + c_2 \frac{1+u^4 + 8 u^2 \log \frac{1-u^2}{u} }{(1+u^2)^2} \ .
\ee
Given this solution, we can attempt a perturbative solution in $\delta \mu = \mu - 4$, $\lambda$ and $\omega$ 
\begin{eqnarray}
a_x^3 &=& (1-u)^{ -\rmi   \omega/4 \pi T} ( a_x + \omega a_{x\omega} + \lambda^2 a_{x \lambda} + \omega \lambda^2 
a_{x \omega \lambda} + \ldots) \ , \\
a_\pm &=& (1-u)^{ -\rmi  \omega/4 \pi T}( a_{\pm} + \omega a_{\pm \omega} + \delta \mu \, a_{\pm \mu}
+ \lambda^2 a_{\pm \lambda} + \ldots ) \ .
\end{eqnarray}
Imposing appropriate boundary conditions, we find
\begin{eqnarray}
a_x &=& 1 \ , \\
a_{x\omega}  &=&-\frac{\rmi}{4} \log \frac{1+u}{1+u^2} \ , \\
a_{x \lambda} &=&  \mu^2 (1-u^2) \ , \\
a_{x \omega \lambda} &=& \frac{\rmi \mu^2}{12 (1+u^2)} \biggl[ -1 + u^2 + 16(1+u^2) \log 2 +(3 u^4 - u^2 - 4) \log(1+u) \nonumber \\
&&
\hspace{20mm} -3(u^4+5 u^2+4) \log(1+u^2) \biggr] \ , 
\end{eqnarray}
\begin{eqnarray}
a_\pm &=& (j_1 \pm \rmi \,  j_2) \frac{u^2}{(1+u^2)^2} \ , \\
a_{\pm \omega} &=& \pm \frac{(j_1 \pm \rmi j_2)(1 \mp \rmi)}{96 (1+u^2)^2}
\biggl[
( 7\pm 4\rmi) - ( 1\pm 4\rmi) u^4 - 6u^2(1 + \log 16)  \nonumber \\
&&
\hspace{10mm}- 4( 1\pm \rmi) u^2 \left( -3 \log(1+u^2) + 11 \log u \pm 3 \rmi \log \frac{1+u}{u} \right) \biggr] \ , \label{logu}\\
a_{\pm \mu} &=& - (j_1 \pm \rmi \, j_2) \frac{(1-u^2)^2 + 8 u^2 \log \frac{1+u^2}{2 u}}{16(1+u^2)^2} \label{logu2}\ .
\end{eqnarray}
The expression for $a_{\pm \lambda}$ is more complicated and we do not list it here. 
We instead include the differential equation that we require $a_{\pm\lambda}$ to satisfy:
\begin{eqnarray}
a''_{\pm\lambda}(u)-\frac{\left(1+3 u^4\right) } {u(1-u^4)} a'_{\pm\lambda}(u)
+\frac{16 }{\left(1+u^2\right)^2}a_{\pm\lambda}(u)+ (j_1 \pm \rmi \,  j_2)  \frac{128 \left(u^4+9 u^2-2\right) u^4}{3 \left(1+u^2\right)^5}=0 \ .
\end{eqnarray}
We can now expand the solutions near $u=0$:
\begin{eqnarray}
a_x^1 &=& a_{10} + a_{11} u + a_{12} u^2  + \ldots \ , \\
a_x^2 &=& a_{20} + a_{21} u + a_{22} u^2 + \ldots \ , \\
a_x^3 &=& a_{30} + a_{32} u^2  + \ldots \ .
\end{eqnarray}
Now we again consider that there are two systems related by a gauge--transformation, and the boundary conditions translate into the following 
three relations that allow us to solve for $j_1$, $j_2$ and $\theta^*$:
\begin{eqnarray}
\label{normalbdry1}
a_{20} &=& 0 \ , \\
\label{normalbdry2}
a_{10} + a_{30} \,\tan \theta   &=& 0 \ ,  \\
\label{normalbdry3}
\left. (a_{12} \, \tan \theta  - a_{32}) \right|_{\omega = 0} &=& 0 \ .
\end{eqnarray}
Finally, using \eqref{sigma}, we obtain a finite DC conductivity:
\begin{eqnarray}
\label{ads5}
\sigma_{DC}(\delta \mu, \lambda)= \frac{1}{2}+\frac{72}{64 (-17+24 \log 2)-9 {\delta \mu\over\lambda ^2} }+\cdots
\end{eqnarray}
It seems that there might be an issue with the non-commutativity in the limits $\delta\mu \to 0$ and $\lambda \to 0$: If we take them in the order $\lambda\to 0$ then $\delta \mu\to 0$, the limit is $\sigma_{DC} \big|_{\delta\mu = \lambda = 0} =\frac{1}{2}$, while in the opposite order $\delta \mu\to 0$ then $\lambda\to 0$ the result is $\sigma_{DC} \big|_{\delta\mu = \lambda = 0} = \frac{1}{2} +\frac{9}{8 (-17 + 24 \log 2)}$. However, looking at the phase diagram Fig.\ \ref{phasediagram} (right), it should be understood that only the first order is allowed if we are to stay in the normal phase as we approach the point $(\lambda=0, \delta\mu=0)$. Fig.~\ref{mueq4} (left) shows a blow up of the phase diagram close to $(\lambda=0, \delta\mu=0)$.
The blue line corresponds to the second order phase transition, located where the denominator in \eqref{ads5} vanishes. 
The green line corresponds to the curve $\delta\mu=0$, or $\mu=4$, and we can see it is outside the normal phase. (The green line curves downward because the Hawking temperature (\ref{HawkingT}) depends on $\lambda$.)
Indeed, the result \eqref{ads5} is only valid in the normal phase where $\frac{\delta \mu}{\lambda^2} < -2.592$.

Fig.\ \ref{mueq4} (right) is a plot of the DC conductivity computed numerically using a method analogous to the one described in section \ref{normal}, and analytically, using the approximation \eqref{ads5}. We see that the agreement is very good. Note that the result (\ref{ads5}) will produce a DC resistivity that is linear in temperature near $T_c$ because $\sigma_{DC}$ has a simple pole at the phase transition temperature.  Although we have little analytic control away from the probe limit in $d=4$, the linear dependence of $\rho_{DC}$ on temperature in the $d=3$ case indicates that $\sigma_{DC}$ continues to have what looks like a simple pole at $T_c$. Finally, notice that in the $AdS_5$ case, we in general encounter an issue with a $\log u$ divergence that makes the charge current regularization scheme dependent. This divergence can be seen for example in \eqref{logu} and \eqref{logu2}. However, this divergence is subleading in the probe limit where we perform our perturbative calculation. Thus, the results obtained in this section do not depend on any regularization scheme.\\

%

\section{Discussion}

Starting from a nonabelian gravitational action, in this paper we showed how to perform a nonabelian gauge transformation to obtain a finite DC conductivity without introducing a lattice or impurities to break the translation symmetry. Close to the first order phase transition, we found a linear temperature dependence of the DC resistivity.  We also obtained an analytic result of finite DC conductivity for the $AdS_5/CFT_4$ case close to the probe limit.

It would be of great interest to get a better physical understanding of the nonabelian gauge transformation and the special isospin rotation angle $\theta^*$ that yields a finite DC conductivity.  
In section 3, we explained how the angle $\theta$ effectively cancels a pole in the DC conductivity for the electric field in the 3 isospin direction against a pole in the conductivity in the 1 and 2 isospin directions.  
In the appendix, we perform an analogous calculation for free fermions transforming under a global SU(2) symmetry.  In that case, we find that for a general representation of SU(2), there will be $1/\omega$ poles in the conductivity in the 1, 2, and 3 isospin directions.  However, the poles all have the same sign, and there is no special angle in isospin space where the poles sum to zero and the DC conductivity remains finite.  It may well be that this angle is a special feature of strong coupling.
The special boundary condition we choose (that cancels the poles) sets the current
$\cal J$ to zero in the DC limit in the transformed frame.  As such, it is reminiscent of the alternate quantization of scalar fields in $AdS_{d+1}$, where one chooses the scaling dimension of the dual field theory operator to be $\Delta < d/2$.

We should emphasize that although we have rendered the DC conductivity finite, our system will still support a persistent energy current $T^{ti}$ in the absence of any sources because of momentum conservation.  In fact, in the context of our model, an electric field in the third isospin direction will act to accelerate the energy density of the system while at the same time producing a steady state charge current (in the third isospin direction), a rather exotic phenomenon as far as we know.  (Assuming that the chemical potential lies in the 13-isospin plane, note that a $F_{tx}^3$ also acts to accelerate a current in the one isospin direction.)

It would be interesting to explore how the non-abelian transformations explored here are affected by the presence of a lattice and impurities as we mentioned earlier. It is also interesting to explore the superconducting $T<T_c$ phase of this system in greater detail in light of our nonabelian gauge transformation.  Finally, to see if more general time and space dependent gauge transformations might lead to more interesting physical effects in these holographic systems is certainly of great interest. 

\section*{Acknowledgments}
We would like to thank Sasha Abanov, Koushik Balasubramanian, Richard Davison, Sean Hartnoll, Kristan Jensen, Subir Sachdev, and William Witczak-Krempa for discussion. We also thank Ra\'{u}l Arias for discussion on the phase diagram in \cite{Arias:2012py}.  This work was supported in part by the National Science
Foundation under Grants No.\ PHY08-44827 and PHY13-16617.  C.~H. thanks the Sloan Foundation for partial support.
The research of R.~V. was partially supported by the FCT-Portugal Grant SFRH/BD/70613/2010.

\appendix

\section{SU(2) Conductivity for Free Fermions}

In this appendix, we calculate the nonabelian current-current Green's function for free fermions transforming under a 
representation of $SU(2)$.  In contrast to the holographic results discussed in the paper, 
we find that there is no direction in isospin space where the DC conductivity remains finite.
We start with the Lagrangian density for free massive fermions coupled to an external $SU(2)$ gauge field:
\be
{\mathcal L} = \rmi \bar \psi \gamma^\mu ( \partial_\mu - \rmi A_\mu^a \tau_a) \psi + M \bar \psi \psi  \ ,
\ee
where as usual $\bar \psi = \psi^\dagger \gamma^t$.  
We choose the explicit representation for the gamma matrices
\be
\gamma^t = \sigma_z \ , \\\
\gamma^x =i  \sigma_x \ , \\\
\gamma^y = i \sigma_y \ ,
\ee
such that $\{ \gamma^\mu , \gamma^\nu \} = -2 \eta^{\mu\nu}$.  We work in mostly plus notation $\eta^{\mu\nu} = (- ++)$.  
For the moment, we will be agnostic about the representation under which the $\psi$ transform. The following approach is valid for any representation.  There is understood to be a hidden index
$\bar \psi \tau_a \psi = \bar \psi_i \tau_a^{ij} \psi_j$.
The $SU(2)$ current is 
\be
J^\mu_a(x) = \bar \psi \gamma^\mu \tau_a \psi \ .
\label{su2J}
\ee
At nonzero temperature and chemical potential, 
we would like to compute the Fourier transform of the retarded Green's function constructed from $J^\mu_a(x)$:
\be
(G_{R})_{ab}^{\mu\nu}(\omega, \vec k) = -\rmi \int  \theta(t)  \left\langle [ J_a^\mu(x), J_b^\nu(0) ] \right\rangle \rme^{-\rmi k \cdot x} \, \rmd^3 x \ .
\ee
As all directions in isospin space are equivalent, we will turn on chemical potential $\mu = A^3_t$ and assume that $\tau_3$ is diagonal.
Since we are interested in the conductivity $\sigma_{xx} = G_R^{xx}(\omega, 0)/(\rmi \omega)$, we will shortly specialize to the $xx$-component of $G_R^{\mu\nu}(\omega,\vec k)$ at zero spatial momentum $\vec k = 0$. 
The most efficient procedure appears to be to compute the Euclidean Green's function using the standard Feynman rules, and then perform the analytic continuation.
Useful references are \cite{Miransky, Sachdev}.
The usual Feynman rules tell us that the Euclidean Green's function is
\be
G^{xx}_{ab} (\rmi \, \Omega_m, {\vec  p}) =2 \pi T \sum_{n=-\infty}^\infty \int \frac{\rmd^2 {\vec k}}{(2\pi)^2}
\tr [ \gamma^x S_i (\Omega_m + \omega_n, {\vec p} + {\vec k}) \gamma^x S_j(\omega_n, {\vec k}) ] \tau_a^{ij} \tau_b^{ji}
\ee
where the fermionic propagator takes the form
\be
S_i(\omega_n , {\vec k}) = \frac{\rmi}{(\rmi \omega_n - \mu_i) \gamma^t + {\vec k} \cdot {\vec \gamma} + M} \ .
\ee
The Matsubara frequencies are $\omega_n = (2n+1) \pi T$ and $\Omega_m = 2 m \pi T$, and a sum on the indices $i$ and $j$ is implied.
The trace over the gamma matrices can be performed and yields
\begin{eqnarray}
\lefteqn{\tr [ \gamma^x S_i (\Omega_m + \omega_n, {\vec p} + {\vec k}) \gamma^x S_j(\omega_n, {\vec k}) ] = }
\nonumber \\
&& 2 \frac{ k_x^2 - k_y^2 +  (\rmi\,\Omega_m + \rmi\, \omega_n-\mu_i) (\rmi\, \omega_n - \mu_j) + k_x p_x - k_y p_y - M^2}
{((\rmi\, \Omega_m + \rmi\, \omega_n - \mu_i)^2 - (\vec p + \vec k)^2 - M^2)(( \rmi\, \omega_n - \mu_j)^2 -  \vec k^2 - M^2)}
\end{eqnarray}
We now specialize to the case where $\vec p = 0$.  We decompose the trace into a sum of three simpler terms:
\begin{eqnarray}
\lefteqn{\tr [ \gamma^1 S_i (\Omega_m + \omega_n, {\vec k}) \gamma^1 S_j(\omega_n, {\vec k}) ] = } \nonumber \\
&& \frac{ 4 k_x^2 - (\rmi\, \Omega_m + \mu_j - \mu_i)^2}{(\vec k^2 + M^2- (\rmi\, \Omega_m + \rmi\, \omega_n - \mu_i)^2  )( \vec k^2 + M^2-( \rmi\, \omega_n - \mu_j)^2  )} \nonumber \\
&& 
-\frac{1}{\vec k^2 + M^2- (\rmi\, \Omega_m +\rmi\, \omega_n - \mu_i)^2  } - \frac{1}{ \vec k^2 + M^2-( \rmi\, \omega_n - \mu_j)^2}  \ .
\end{eqnarray}
Before performing the sum over Matsubara frequencies, it is convenient to integrate by parts the last two terms above with respect to $k_x$
\cite{Sachdev}.
It is then manifest that $G^{xx}_{ij}(\rmi\, \Omega_m, 0)$ vanishes in the limit $T, \mu_i, \mu_j, \Omega_m \to 0$.  
We find that
\begin{eqnarray}
G^{xx}_{ij} (\rmi\,\Omega_m, 0) &=& 2 \pi T \sum_{n=-\infty}^\infty \int \frac{\rmd^2 {\vec k}}{(2\pi)^2}
\biggl[
\frac{ 4 k_x^2 - (\rmi\, \Omega_m + \mu_j - \mu_i)^2}{\left(\vec k^2 + M^2- (\rmi\, \Omega_m + \rmi\, \omega_n - \mu_i)^2  \right)\left( \vec k^2 + M^2-( \rmi\, \omega_n - \mu_j)^2  \right)} \nonumber \\
&& 
\hskip 0.2in 
-\frac{2 k_x^2}{\left(\vec k^2 + M^2- (\rmi\, \Omega_m + \rmi\, \omega_n - \mu_i)^2 \right)^2 } - \frac{2 k_x^2}{\left( \vec k^2 + M^2-( \rmi\, \omega_n - \mu_j)^2\right)^2} 
\biggr] \ .
\end{eqnarray}
Performing the sum over the Matsubara frequencies now yields
\begin{align}
G^{xx}_{ij} (\rmi\,\Omega_m, 0) =
 \frac{\pi}{2}  \int \frac{\rmd^2 {\vec k}}{(2\pi)^2}
& \biggl[
\frac{ k_x^2}{2T \epsilon^2 } \left( \sech^2 \frac{\epsilon+\mu_i}{2T} + \sech^2 \frac{\epsilon-\mu_i}{2T} 
+ \sech^2 \frac{\epsilon+\mu_j}{2T} + \sech^2 \frac{\epsilon-\mu_j}{2T} \right)
\nonumber \\
&
- \frac{k_x^2 ( 4 \epsilon^2 + 2 \epsilon \tilde \omega + \tilde \omega^2) - \epsilon^2 \tilde \omega^2}
{\epsilon^3 \tilde \omega(2 \epsilon + \tilde \omega)}
\left(\tanh \frac{\epsilon +\mu_j}{2 T} + \tanh \frac{\epsilon - \mu_i}{2 T} \right) \nonumber \\
&
+ \frac{k_x^2 ( 4 \epsilon^2 - 2 \epsilon \tilde \omega + \tilde \omega^2) - \epsilon^2 \tilde \omega^2}
{\epsilon^3 \tilde \omega(2 \epsilon - \tilde \omega)}
\left(\tanh \frac{\epsilon -\mu_j}{2 T} + \tanh \frac{\epsilon + \mu_i}{2 T} \right) 
\biggr]
\end{align}
where we have introduced the shifted frequency variable $\tilde \omega = \rmi\, \Omega_m - \mu_i + \mu_j$. 
Note that we have used the fact that $\sech^2$ and tanh are periodic under shifts by $\rmi\, \Omega_m$.  
We now change variables to polar coordinates. Integrating over the angle allows us to replace $k_x^2$ with $\vec k^2 / 2$.  We then change coordinates again to $\epsilon = \sqrt{\vec k^2 + M^2}$, and the Green's function becomes
\begin{align}
G^{xx}_{ij} (\rmi\,\Omega_m, 0) =
\frac{1}{8}  \int_{|M|}^\infty \epsilon &  \, \rmd\epsilon
 \biggl[
\frac{ \epsilon^2-M^2}{2T \epsilon^2 } \left( \sech^2 \frac{\epsilon+\mu_i}{2T} + \sech^2 \frac{\epsilon-\mu_i}{2T} 
+ \sech^2 \frac{\epsilon+\mu_j}{2T} + \sech^2 \frac{\epsilon-\mu_j}{2T} \right)
\nonumber \\
- & \frac{(\epsilon^2-M^2) ( 4 \epsilon^2 + 2 \epsilon \tilde \omega + \tilde \omega^2) - 2\epsilon^2 \tilde \omega^2}
{\epsilon^3 \tilde \omega(2 \epsilon + \tilde \omega)}
\left(\tanh \frac{\epsilon +\mu_j}{2 T} + \tanh \frac{\epsilon - \mu_i}{2 T} \right) \nonumber \\
+ & \frac{(\epsilon^2-M^2) ( 4 \epsilon^2 - 2 \epsilon \tilde \omega + \tilde \omega^2) - 2\epsilon^2 \tilde \omega^2}
{\epsilon^3 \tilde \omega(2 \epsilon - \tilde \omega)}
\left(\tanh \frac{\epsilon -\mu_j}{2 T} + \tanh \frac{\epsilon + \mu_i}{2 T} \right) 
\biggr] \ .
\end{align}
To get the retarded Green's function we make the substitution $\rmi\, \Omega_m = \omega + \rmi\, \delta$ where $\delta >0$ is infinitesimal.
Following ref.\ \cite{Sachdev}, we divide the Green's function into a quasiparticle contribution and a coherent contribution:
\be
(G_R)^{xx}_{ij} (\omega,0) = G^{\rm qp}_{ij} + G^{\rm coh}_{ij}(\omega + \rmi\, \delta,0) \ .
\ee
where
\[
G^{\rm qp}_{ij} \equiv \frac{1}{8} \int_{|M|}^\infty  \rmd\epsilon
\frac{ \epsilon^2-M^2}{2T \epsilon } \left( \sech^2 \frac{\epsilon+\mu_i}{2T} + \sech^2 \frac{\epsilon-\mu_i}{2T} 
+ \sech^2 \frac{\epsilon+\mu_j}{2T} + \sech^2 \frac{\epsilon-\mu_j}{2T} \right)
\]
and
\begin{eqnarray*}
G^{\rm coh}_{ij}(\omega ,0) &\equiv & \frac{1}{8} \int_{|M|}^\infty \rmd\epsilon
\biggl[
- \frac{(\epsilon^2-M^2) ( 4 \epsilon^2 + 2 \epsilon \tilde \omega + \tilde \omega^2) - 2\epsilon^2 \tilde \omega^2}
{\epsilon^2 \tilde \omega(2 \epsilon + \tilde \omega)}
\left(\tanh \frac{\epsilon +\mu_j}{2 T} + \tanh \frac{\epsilon - \mu_i}{2 T} \right) \nonumber \\
&&
+\frac{(\epsilon^2-M^2) ( 4 \epsilon^2 - 2 \epsilon \tilde \omega + \tilde \omega^2) - 2\epsilon^2 \tilde \omega^2}
{\epsilon^2 \tilde \omega(2 \epsilon - \tilde \omega)}
\left(\tanh \frac{\epsilon -\mu_j}{2 T} + \tanh \frac{\epsilon + \mu_i}{2 T} \right) 
\biggr] \ .
\end{eqnarray*}

As we divide $G^{\rm qp}_{ij}$ by $\rmi\,\omega-\delta$ to get the quasiparticle contribution to the conductivity, this term will yield a Dirac delta function $\delta (\omega)$ contribution to $\operatorname{Re}(\sigma^{xx}(\omega))$.
While we have not succeeded in obtaining an analytic expression for the ${\cal O}(M^2)$ term, the leading terms in a small $M$ expansion
evaluate to
\begin{eqnarray}
G^{\rm qp}_{ij} (\omega, 0) &=& \frac{T}{4}  \log \left( \left(1 + \rme^{(M-\mu_i)/T}\right)\left(1+\rme^{(M+\mu_i)/T}\right)\left(1+\rme^{(M+\mu_j)/T}\right)\left(1+\rme^{(M-\mu_j)/T}\right) \right)
 \\
&& 
-\frac{M}{8} \left( 4 + \tanh \frac{M-\mu_i}{2T} + \tanh \frac{M+\mu_i}{2T} + \tanh \frac{M-\mu_j}{2T} + \tanh \frac{M+\mu_j}{2 T} \right) 
+ {\cal O}(M^2) \ .
\nonumber
\end{eqnarray}
For $M=0$ and $\mu_i = \mu_j = 0$, we find  
\be
G^{\rm qp}_{ij} =  T \log 2 \ .
\ee
The delta function here comes from ballistic transport of the thermally excited fermions.  As there are equal numbers of fermions and anti-fermions at zero chemical potential, we expect that this delta function is broadened by interactions \cite{Sachdev}.
For $M=0$ and $|\mu_j|$, $|\mu_i| \gg T$, we get instead
\be
\label{zeroT}
G^{\rm qp}_{ij} = \frac{1}{4} \left(| \mu_i| + |\mu_j| + T \, {\cal O}(\rme^{-\mu_i/T}, \rme^{-\mu_j/T}) \right) \ .
\ee
This delta function is proportional to the density and should not be broadened.  
The delta function is physical and corresponds 
to the fact that a charged
material will accelerate in response to an electric field. 

Regarding the coherent contribution, we can evaluate the imaginary part of the self-energy through a contour integration.  The result is
\begin{eqnarray}
\operatorname{Im} (G^{\rm coh}_{ij}(\omega, 0)) &=&\frac{ \pi }{8} 
 \Theta(\tilde \omega^2 - 4M^2) \frac{\tilde \omega}{2}  \left( 1 + \frac{4M^2}{\tilde \omega^2} \right)
\left( \tanh \frac{ \tilde \omega + 2 \mu_i}{4 T} + \tanh \frac{ \tilde \omega - 2 \mu_j}{4 T} \right) \ .
\end{eqnarray}
(The jumps in the derivatives of this function at $\tilde \omega^2 = 4 M^2$ should be smoothed by interactions \cite{Sachdev}.)
In the case $\mu_i \neq \mu_j$, it is possible for the Heaviside theta function to evaluate to one at $\omega = 0$.  However, the coefficient of the Heaviside theta vanishes at $\omega=0$ and will not contribute a $\delta(\omega)$ to $\operatorname{Re}(\sigma^{xx}(\omega))$.
For $\mu_i = \mu_j$, the coefficient is nonzero in the limit $\omega \to 0$.  However, the Heaviside theta itself will now evaluate to zero.  In short, the coherent piece will not contribute a $\delta(\omega)$ to the real part of the conductivity even when $\mu_i$ and $\mu_j$ are nonzero.  The $\delta(\omega)$ contribution comes only from $G^{\rm qp}_{ij}$.  

Let us now consider $\psi$ in an arbitrary representation of $SU(2)$.  
Indexing using the angular momentum in the $z$-direction, for a representation of spin-$\ell$, the $\tau$ matrices can be written in the usual way
\begin{eqnarray}
\tau_3 &=& m_1 \delta_{m_1, m_2} \ , \\
\tau_{\pm} &=& \sqrt{(\ell \mp m_1)(\ell \pm m_1 + 1)} \, \,  \delta_{m_2 , m_1 \pm 1} \ ,
\end{eqnarray}
where $m_1, m_2 = -\ell, -\ell+1, \ldots, \ell$.  We also have the relations $\tau_1 = (\tau_+ + \tau_-)/2$ and $\tau_2 = (\tau_+ - \tau_-)/2\rmi$.

We are interested in the $\delta(\omega)$ contribution to the real part of the conductivity in the limit $T = 0$.  We will use the result (\ref{zeroT}). From the Lagrangian, if we take $A_t^3 = \mu>0$, then the chemical potentials are read from the diagonal elements of $\tau_3$: $\mu_i = (\tau_{3})_{ii} \, \mu$.
As we can verify example by example, for spin-$\ell$, the Green's functions take the form\footnote{%
$\lfloor x \rfloor$ is the integer part of $x$.}
\begin{eqnarray}
\operatorname{Im}(G_{33}^{xx}(\omega,0)) &=& \frac{\mu}{8} \left \lfloor \frac{(2\ell+1)^2 - 1}{8} \right \rfloor  + {\cal O}(\omega^2)\ , \\
\operatorname{Im}(G_{11}^{xx}(\omega,0)) = \operatorname{Im}(G_{22}^{xx}(\omega,0)) &=& \frac{\mu}{8} \left \lfloor \frac{(2\ell+1)^2}{16} \right \rfloor  + {\cal O}(\omega^2)\ .
\end{eqnarray}
These Green's functions lead to $\delta(\omega)$ contributions to $\operatorname{Re}(\sigma^{xx}_{ab}(\omega))$.
(The mixed components $G_{ab}^{xx}$ with $a \neq b$ will vanish.)

The way to see there is no special direction in isospin space where the $\delta (\omega)$ contribution vanishes is as follows. 
Continuing to work in a frame where the chemical potential points in the 3-isospin direction (the untransformed or untilded frame in the main text), we apply an electric field in the 13-isospin plane $\vec E = (E^1, E^3)$.  
This electric field will produce a delta function current response of the form $\vec J \sim (\operatorname{Im}(G_{11}^{xx}(0)) E_1, \operatorname{Im}(G_{33}^{xx}(0)) E_3)$.
We then look for a special angle $\tan \theta^* = - E^1/E^3$ such that the dot product
\bea
\vec E \cdot \vec J= 0 
\eea  
vanishes. 
%
The observation is that, in the holographic case, for chemical potential in the 3-isospin direction, $G_{11}$ and $G_{22}$ had opposite sign from $G_{33}$, while the free theory calculation gives $G_{11}$, $G_{22}$ and $G_{33}$ all with the same sign. The pole then cannot be cancelled in a free theory via a special rotation, unlike the holographic result considered in the body of the paper.\footnote{%
See discussion below \eqref{fluctnormal2} for the AdS/CFT case.}  
While it is not obvious to us how precisely interactions alter this story, we tentatively conclude that the interactions between electrons may play a crucial role in explaining our AdS/CFT results regarding the finite DC resistivity with a linear temperature dependence.




\end{document}